\documentclass[preprint,review,authoryear,12pt,3p]{elsarticle}

\usepackage{graphicx,subfigure}
\usepackage{ctable}
\usepackage{amssymb}
\usepackage{amsmath}
\usepackage{bbm}
\usepackage{wrapfig}
\usepackage{cases}
\usepackage[normalem]{ulem}
\newcommand{\ud}{\mathrm{d}}

\usepackage[english]{babel}
\usepackage{amsmath,amsthm}
\usepackage{amsfonts}
\usepackage{enumerate}
\theoremstyle{definition}

\theoremstyle{remark}

\numberwithin{equation}{section}
\usepackage{graphicx,subfigure}
\usepackage{ctable}
\usepackage{amsmath} % If you want to use AMS-LaTex,comment out these
\usepackage{amssymb}
\usepackage{epsfig,amsthm}
\usepackage{float}
\usepackage{epstopdf}

\usepackage{amsmath}
\usepackage{epsfig,amsthm}
\usepackage{amssymb,amsfonts}
\usepackage{dsfont}
\usepackage{graphicx}
\usepackage{subfigure}
\usepackage{indentfirst}
\usepackage{mathrsfs}
\usepackage{color}

\newcommand{\D}{\mathrm{d}}

\journal{}

\begin{document}

\begin{frontmatter}



\title{Dislocation climb models from atomistic scheme to dislocation dynamics}%


 \author[hkust]{Xiaohua Niu}
\author[hkust]{Tao Luo}
\author[duke]{Jianfeng Lu}
 \author[hkust]{Yang Xiang\corref{cor1}}
\ead{maxiang@ust.hk}

\address[hkust]{Department of Mathematics, The Hong Kong University of Science and Technology, Clear Water Bay, Kowloon, Hong Kong}

\cortext[cor1]{Corresponding author}

\address[duke]{Department of Mathematics, Department of Physics and Department of Chemistry, Duke University, Durham, North Carolina 27708, USA}

\begin{abstract}

We develop a mesoscopic dislocation dynamics model for vacancy-assisted dislocation climb by upscalings from a stochastic model on the atomistic scale. Our models  incorporate microscopic mechanisms of (i) bulk diffusion of vacancies,
               (ii) vacancy exchange dynamics between bulk and dislocation core,
                (iii) vacancy pipe diffusion along the dislocation core, and
               (iv) vacancy attachment-detachment kinetics at jogs leading  to the motion of jogs.
 Our mesoscopic model consists of the vacancy bulk diffusion equation and a dislocation climb velocity formula. The effects of these microscopic mechanisms are incorporated  by a  Robin boundary condition near the dislocations for the bulk diffusion equation
  and a new contribution in the dislocation climb velocity due to vacancy pipe diffusion driven by the stress variation along the dislocation.
  Our climb formulation is able to quantitatively describe the translation of prismatic loops at low temperatures when the bulk diffusion is negligible.
Using this new formulation, we derive  analytical formulas for the climb velocity of a straight edge dislocation and a prismatic circular loop. Our dislocation climb formulation can be implemented in  dislocation dynamics simulations to incorporate all the above four microscopic mechanisms of dislocation climb.

\end{abstract}


\end{frontmatter}

%\author{The American Physical Society}%
%\email[REVTeX Support: ]{revtex@aps.org}
%\affiliation{1 Research Road, Ridge, NY 11961}
%\date{August 10, 2010}%

\section{Introduction}
 Dislocation climb  plays significant roles in the plastic deformation of crystals at high temperatures~\citep{Hirth-Lothe}. In the climb process, dislocations change slip planes by absorbing and/or emitting vacancies. This process is driven by the climb component of the Peach-Koehler force on the dislocations, and is coupled with vacancy diffusion in the material.

 Mobility law, which is a proportional dependency between the dislocation climb velocity and the climb force, was commonly used in the early discrete dislocation dynamics simulations \citep{Ghoniem,Xiang20035499,Xiang2006,Arsenlis,Mordehai,Bako,Keralavarma,Ayas2012,Danas2013,Raabe2013}.
{  \citet{Mordehai} incorporated a climb velocity-force relation derived for straight dislocations considering the cases where the far field condition of the vacancy concentration is the same or different from its equilibrium value (e.g., due to irradiation).
\citet{Bako} proposed a climb velocity-force relation based on a single straight dislocation and considered the interaction between different loops through  the change of the far field vacancy concentration which is associated with the condensation of vacancies into loops during coarsening processes.
\citet{Keralavarma} incorporated the connection among the far field vacancy concentration  for each dislocation by solving a vacancy diffusion equation with a source term depending on the dislocation density averaged over a finite volume element, in addition to a local mobility law-like  {\it ansatz} in their two-dimensional dislocation dynamics simulations.}
These mobility-law type dislocation climb velocity formulas were mainly based on the analytical expression of the climb velocity for a single, straight edge dislocation~\citep{Hirth-Lothe}.
   Recently, \citet*{Gu2015319}  derived a  formulation that gives the climb velocity of curved dislocations and multiple dislocations in three-dimensions using the Green's function representation of the solution of vacancy diffusion over the bulk of the material.

    Most of these works on diffusion-controlled dislocation climb dynamics  were  based on vacancy diffusion in the bulk, which are believed to be the dominant control mechanism because  vacancy diffusion at dislocation cores (pipe diffusion) are commonly orders of magnitude faster than that in the bulk~\citep{Hirth-Lothe,Hoagland1998,Fang2000,Mishin2009}. As a result, the vacancy concentrations along dislocation cores were assumed to be in equilibrium, and the Dirichlet boundary condition was usually adopted for the vacancy diffusion in the bulk, meaning that dislocations were always considered as perfect sources or sinks of vacancies.
    However, real dislocations may not necessarily be perfect sources/sinks,
     and both pipe diffusion and the structure of jogs on dislocations play essential roles in the dislocation climb process~\citep{Hirth-Lothe,Balluffi}. In fact,  the  absorption and  emission of vacancies  at the jogs on the dislocations result in the motion of these jogs, leading to climb of the dislocations.

      Only a few attempts  have been made in the literature to incorporate some details of pipe diffusion and jog dynamics in the dislocation climb models. \citet{Gao20111055} incorporated vacancy pipe diffusion in the three dimensional discrete dislocation dynamics model, while the bulk diffusion of vacancies was neglected in their model. { \citet{Danas2013} identified two limit cases associated with the vacancy pipe diffusion. In addition to the diffusion-limited case in the classical dislocation climb theory~\citep{Hirth-Lothe}, they also discussed the sink-limited case where the dislocation climb process is dominated by vacancy absorption to the isolated jogs located far apart from each other along the dislocation. They derived  mobility law-like climb velocity formula for both cases.  Their formulation was adopted by \citet{Ayas2012} in the simulations of tensile response of passivated films with climb-assisted dislocation glide.
    \citet{Ayas2014113} developed a  two-dimensional discrete dislocation plasticity framework coupled to vacancy diffusion. In these works,  the vacancy diffusion boundary value
problem with climbing dislocations as line sources/sinks of vacancies
was also solved by a superposition of the fields of the line sources/sinks in an
infinite medium (obtained by analytical formula) and a complementary non-singular solution that enforces the boundary conditions (solved by finite element method).  For the fields in an infinite medium, \citet{Ayas2014113}  summed up the contributions from all individual dislocations based on the analytical solution formula of the vacancy bulk diffusion problem in the infinite medium with a single edge dislocation as a source/sink, using a Robin boundary condition for the vacancy concentration near the dislocation to account for the effect of vacancy pipe diffusion.}  Although they qualitatively explained the dependence of the coefficient in the Robin boundary condition, its quantitative expression was not given.
    \citet{Pierre-Antoine} proposed a multiscale approach to model  dislocation climb at mesoscopic length and time scales. They analyzed the climb of a single, nominally straight edge dislocation at a nanoscopic scale by solving both vacancy pipe diffusion and bulk diffusion equations. They obtained the dislocation climb velocity of a single edge dislocation, and further incorporated it in a mesoscopic phase-field model of dislocation climb.

 In this paper, we develop a mesoscopic dislocation dynamics model for vacancy-assisted dislocation climb by upscalings from a stochastic model on the atomistic scale. Our models
      incorporate the following microscopic mechanisms that are involved in the dislocation climb process:
      \begin{enumerate}[(i)]
      \item
      Bulk diffusion of vacancies;
      \item
      Vacancy exchange dynamics between bulk and dislocation core;
       \item
      Vacancy pipe diffusion along the dislocation core;
      \item
      Vacancy attachment-detachment kinetics at jogs leading  to the motion of jogs.
      \end{enumerate}

   We first develop a stochastic model on the microscopic scale for dislocation climb incorporating the above microscopic mechanisms. This microscopic model is based on a simple cubic lattice.
       By upscaling in space and time from the microscopic model, we obtain the jog dynamics model. This model consists of (1) a vacancy pipe diffusion equation on the jogged dislocation line with boundary condition at each jog depending on the vacancy attachment-detachment kinetics there, (2) a vacancy bulk diffusion equation with boundary conditions on the dislocation depending on the vacancy exchange between the bulk and the dislocation core, and (3) a jog velocity formula for the motion of jogs along the dislocation which comes from the inward flux of vacancies and leads to the climb of the dislocation. { Generalizations are made to eliminating the simple-cubic-lattice dependent features in the derived jog dynamics model, so that it can apply to general lattice structures as the classical dislocation climb theory \citep{Hirth-Lothe}.}

       By further upscaling in space and time from the jog dynamics model to the mesoscopic scale, we derive the  dislocation dynamics model for vacancy-assisted dislocation climb,   based on the fact that vacancy pipe diffusion is commonly orders of magnitude faster than that in the bulk~\citep{Hirth-Lothe,Hoagland1998,Fang2000,Mishin2009,Gu2016}. This mesoscopic model consists of the vacancy bulk diffusion equation and a dislocation climb velocity formula.
       The effects of vacancy pipe diffusion and absorption/emission on jogged  dislocations and the vacancy exchange between dislocation core and the bulk are incorporated in this mesoscopic model by a  Robin boundary condition for the bulk diffusion equation and a new contribution in the dislocation climb velocity due to the stress variation along the dislocation. Especially, the new contribution in the climb velocity depends only on the pipe diffusion. Our new climb formulation can be employed to quantitatively describe the translation of a prismatic loop at low temperatures when the bulk diffusion is negligible. Although this "self-climb" mechanism was proposed and observed in experiments  quite a long time ago \citep{Johnson1960,Kroupa1961,Hirth-Lothe}, a widely-accepted quantitative theory for it is still lacking.
        One effort  \citep{Kroupa1961} is to explain it qualitatively by the elastic interaction between a straight edge dislocation and a circular prismatic loop, without considering the details of pipe diffusion. Some other available theories used mobility law of the entire prismatic loop to account for the effect of self climb, and the variation of the interaction energy of the entire loop with the stress field as the driving force  \citep{Turnbull1970,Dudarev2016}. Our formulation is able to give the climb velocity quantitatively by vacancy pipe diffusion under stress variation at any point on arbitrarily curved dislocations.

       Using this new formulation, we also derive analytical formulas for the climb velocity of a straight edge dislocation and a circular prismatic loop, incorporating all the above four microscopic mechanisms. In the limit cases of fast exchange of vacancies between the dislocation core and the bulk and small stress variation, our formulation is reduced to the classical results. Our formulation can be implemented in three dimensional dislocation dynamics simulations, for example, efficiently using the Green's function or the boundary integral equation method as in \citet{Gu2015319}. This is being explored and results will be reported elsewhere.

        The rest of the paper is organized as follows. In Sec.~\ref{sec:II}, we formulate the discrete stochastic scheme for dislocation climb on the atomistic level, which accounts for all the four microscopic mechanisms above. In Sec.~\ref{sec:III}, by upscaling the microscopic model, we  derive the jog dynamics model that consists of a pipe diffusion equation of vacancies, a bulk diffusion equation of vacancies, and the velocity of a jog. In Sec.~\ref{sec:IV}, by upscaling of the jog dynamics model, we obtain the  dislocation dynamics model for vacancy-assisted dislocation climb, which consists of a diffusion equation for vacancy bulk diffusion and a dislocation climb velocity formula. In Sec.~\ref{sec:V}, we use the new formulation to calculate the climb velocity of a straight edge dislocation and a circular prismatic loop, and explain quantitatively the translation of prismatic loops at low temperatures.

  \section{The microscopic scheme}\label{sec:II}

  In this section, we present a stochastic model on the microscopic scale for dislocation climb, incorporating the  four microscopic mechanisms listed in the introduction section. Similar microscopic model for the one-dimensional adatom diffusion and step motion in epitaxial growth and its step dynamics limit were considered by \citet{Lu2015}.

  \begin{figure}[htbp]
  \centering
   \subfigure[]
         {\includegraphics[width=10cm]{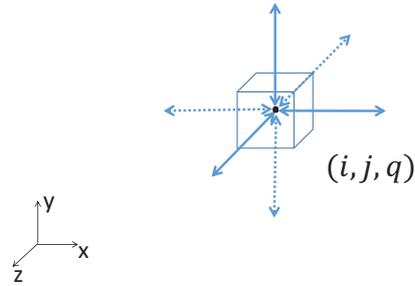} \label{cubic}  }
  \subfigure[]
         { \includegraphics[width=12cm]{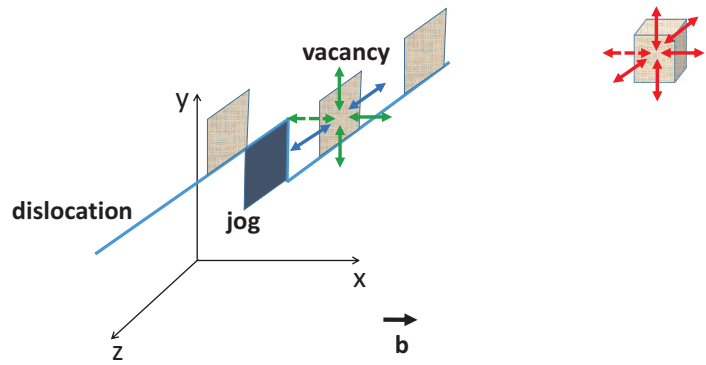}\label{dislocation}}
  \caption{(a) Each vacancy occupies a lattice site in the simple cubic lattice in three dimensions denoted by a triple of integers $(i,j,q)$, and can hop to one of its nearest neighbors in the $x$, $y$, or $z$ direction. Conversely, vacancies can also hop from its nearest neighbor sites to itself in these three directions. (b) A dislocation with a jog in the simple cubic lattice. The dislocation is a line of lattice sites, and a jog occupies a lattice site. For convenience of illustration, the dislocation is drawn as a line, and the site of a vacancy or a jog on the dislocation is plotted as a square in the $yz$ plane. The vacancies could hop between bulk sites, sites on the dislocation, and between a site on the dislocation and a site in the bulk. }
\end{figure}

    \subsection{Assumptions}

 We consider a dislocation in a simple cubic lattice in three dimensions, with lattice constant being the length of Burgers vector $b$ of the dislocation.
    A cubic site in the lattice  is denoted by  $(i, j, q)$ where $i, j, q$ are integers, meaning that the center of the cube is located at $(ib, jb, qb)$ in the three-dimensional space, see Fig.~\ref{cubic}. Each vacancy is of the size of a lattice site, and can hop to one of its nearest neighbors. The vacancies could hop between bulk sites, sites on the dislocation, and between a site on the dislocation and a site in the bulk, see Fig.~\ref{dislocation}.
The dislocation is a line of lattice sites, which is nominally parallel to the $z$ axis and lies in the $yz$ plane, with jogs on it of height $b$ and in the $y$ direction, see Fig.~\ref{dislocation}. This means that the dislocation core size is $b$. The Burgers vector $\mathbf b$ of the dislocation is in the $+x$ direction.
    Each jog on the dislocation occupies a lattice site and serves as a sink or source of vacancies.
    When a jog absorbs or emits a vacancy, it moves forward or backward along the dislocation core by a lattice site. We say a jog is upward meaning that the dislocation goes to a higher slip plane across the jog in the $+z$ direction, e.g. the jog shown in Fig.~\ref{dislocation}; otherwise, we say the jog is downward.

    We apply the following main assumptions for the hoping of vacancies:
      \begin{enumerate}[(1)]
      \item
    A vacancy is only allowed to hop to one of its nearest neighbors in the $x$, $y$, or $z$ direction as shown in Fig.~\ref{cubic}. Nucleation of vacancy clusters is neglected. The elastic interactions among vacancies and between the vacancies and the dislocation are also neglected.

     \item
     The jogs on the dislocations are well-separated. We only consider the pre-existing jogs and neglect their nucleation.

    \item
     In the bulk away from the dislocation, a vacancy can hop from a lattice site to an adjacent site with  rate $\Gamma_v=\Gamma_v^0e^{-\frac{E^v}{kT}}$, where $k$ is Boltzmann's constant,  $T$ is the  temperature, $E^v$ is the activation energy for vacancy diffusion in the bulk,  and $\Gamma_v^0$ is a temperature-independent parameter  related to the Debye frequency $\nu$, vacancy vibrational entropy and the lattice constant, etc. \citep{Balluffi}. (Note that $E^v$, $\Gamma_v^0$, and values of parameters in other assumptions can be obtained for example by atomistic calculations, e.g.~\citet{Hoagland1998,Fang2000,Mishin2009,Yip2009,Yip2010,YZWang2012,Dudarev2016}.)

    \item
    Along the dislocation core, a vacancy can hop from a lattice site to an adjacent site in the dislocation core with rate $\Gamma_c=\Gamma_c^0e^{-\frac{E^c}{kT}}$. As in the hopping of vacancies in the bulk, $E^c$ is the activation energy for vacancy pipe diffusion, and $\Gamma_c^0$ is a temperature-independent parameter. Note that pipe diffusion of vacancies is commonly orders of magnitude faster than that in the bulk~\citep{Hirth-Lothe,Hoagland1998,Fang2000,Mishin2009}.

    \item
     At a lattice site adjacent to the dislocation core, vacancies can hop not only to the five adjacent bulk lattice sites, but also into the dislocation core with rate $\Gamma_v \phi_v=\Gamma^0_{v-c} e^{-\frac{E^{v-c}}{kT}} $. Here $E^{v-c}$ is the energy barrier for the vacancy hopping from the bulk to the dislocation core, $\Gamma^0_{v-c}$ is a temperature-independent parameter, and $\phi_v$ is a dimensionless parameter indicating the difference between the hopping rate from a bulk site to a neighboring bulk site and that from a bulk site to a neighboring site in the dislocation core.
     On the other hand, the vacancy hopping rate from a site in the dislocation  core to the bulk  is $\Gamma_v \phi_vk_v=\Gamma^0_{c-v} e^{-\frac{E^{c-v}}{kT}} $, where $E^{c-v}$ is the energy barrier for the vacancy hopping from dislocation core to the bulk, $\Gamma^0_{c-v}$ is a temperature-independent parameter, and $k_v$ is a dimensionless parameter indicating the difference between the hopping rates out of and into the dislocation core.

     \item
      At the lattice site of a jog, we assume that vacancies  may hop to one of its two adjacent sites along the dislocation core with the rate $\Gamma_c$  as given in assumption (4). We also assume that the rates in assumption (5)  apply to the exchange of vacancies between the  jog site and the bulk. That is,  vacancies may hop into the jog site from a neighboring site in the bulk with the rate $\Gamma_v \phi_v$,  and may hop from the jog site to a neighboring site in the bulk with the rate $\Gamma_v \phi_vk_v$. The jog moves forward or backward when it absorb or emits a vacancy.

\item
      Near the lattice site of a jog, vacancies reach  thermodynamic equilibrium where the osmotic force due to vacancy diffusion (mainly from the pipe diffusion) balances the climb Peach-Koehler force on the dislocation (by the same argument as that of the total free energy equilibrium  in the classical theory where any point on the dislocation is considered as a sink of vacancies, e.g. \citet{Hirth-Lothe,Gu2015319}).  This process can be assumed to be instantaneous because it is much faster than the vacancy concentration equilibration by pipe and bulk diffusions~\citep{Hirth-Lothe}. {The equilibrium vacancy concentration near a jog site is given by  $c_J=c_0^ce^{-\frac{f_{\rm cl}\Omega}{bkT}}$, where $c_0^c=e^{-\frac{E^{f}_c}{kT}}$ is the reference equilibrium vacancy concentration in the dislocation core region with $E^{f}_c$ being the vacancy formation energy within the dislocation core,} $f_{\rm cl}$ is the climb component of the Peach-Koehler force on the dislocation, and $\Omega$ is the volume of an atom. Recall that the full Peach-Koehler force is $\mathbf f=(\pmb\sigma\cdot \mathbf b)\times \pmb\xi$, where $\pmb\sigma$ is the stress tensor and $\pmb\xi$ is the dislocation line direction.  In our atomistic scheme, we set the probability of finding a vacancy at a jog site to be $c_J=c_0^ce^{-\frac{f_{\rm cl}\Omega}{bkT}}$.

    \end{enumerate}

{Note that the reference equilibrium vacancy concentration in the dislocation core region $c_0^c$ is related to the reference equilibrium vacancy concentration in the bulk $c_0$ by $k_vc_0^c=c_0$, due to the detailed balance for the vacancy hopping between the dislocation core and the bulk, i.e. $\Gamma_v \phi_vk_vc_0^c=\Gamma_v \phi_vc_0$. Recall that the reference equilibrium vacancy concentration in the bulk $c_0=e^{-\frac{E^{f}_v}{kT}}$, where $E^{f}_v$ is the vacancy formation energy in the bulk. Similar parameters have been used by \citet{Pierre-Antoine} in their  multiscale theory at coarser level based on dislocation jog dynamics.}

\subsection{The microscopic scheme}

 We consider the system at discrete time $t=t_n$, $n=0,1,2,\cdots$, with a constant and sufficiently small time step $\tau=t_{n+1}-t_n$.
  After each discrete time $t_n$, a jog may move by one lattice site in a stochastic way with a probability that depends on the hopping of vacancies into and out of the site of the jog, and the probability of finding a vacancy at each lattice site
     is updated to describe the averaged effect over the time period $[t_n,t_{n+1}]$.
{ We let variables $c_{i,j,q}^{v,n}$ and $c_{i,j,q}^{c,n}$ be the probabilities of finding a vacancy  at the lattice site $(i, j, q)$ at the time $t_n$ when it is a bulk site and a site in the dislocation core, respectively. Note that the macroscopic counterpart of the probability of finding a vacancy on a lattice site is the vacancy concentration (density) at a physical location, and hence the notation $c$ is used here.} Following the assumptions in the previous subsection, we have the following formulas for the distribution of vacancies and  motion of jogs.

 \begin{figure}[htbp]
  \centering
  \subfigure[]
       {\includegraphics[width=8cm]{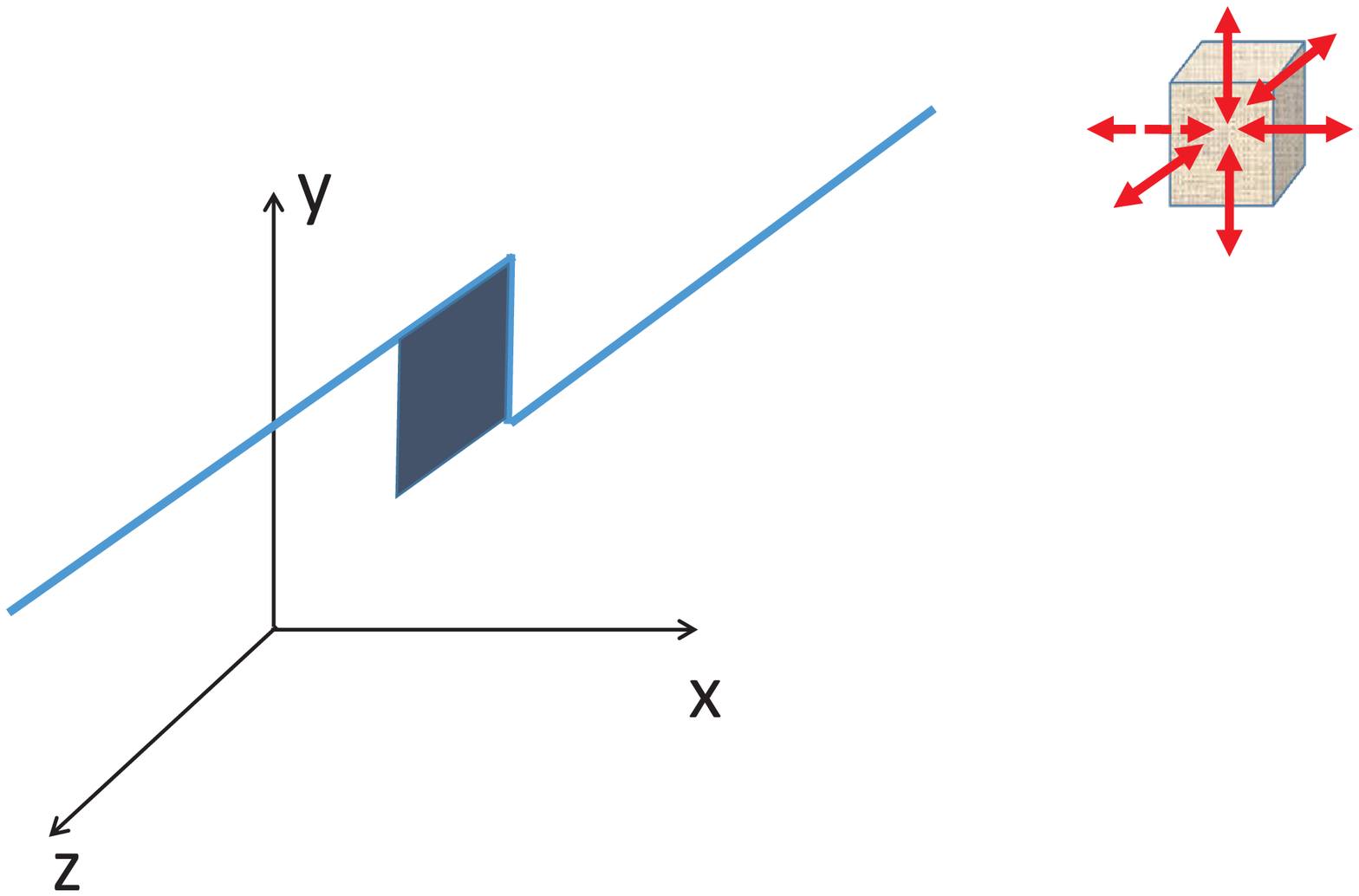}\label{bulk}}
  \subfigure[]
       {\includegraphics[width=8cm]{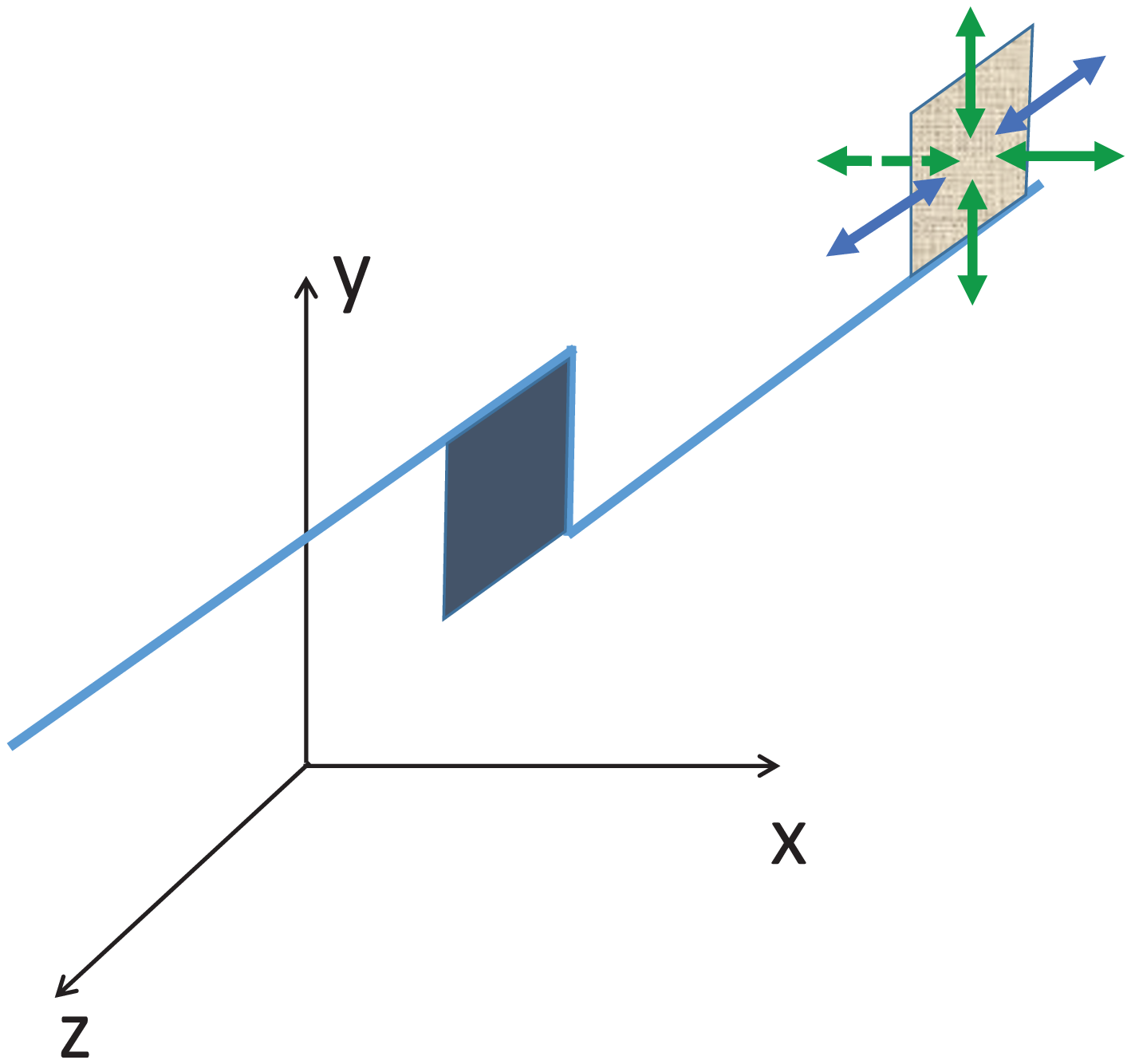} \label{pipe}}
  \subfigure[]
       {\includegraphics[width=8cm]{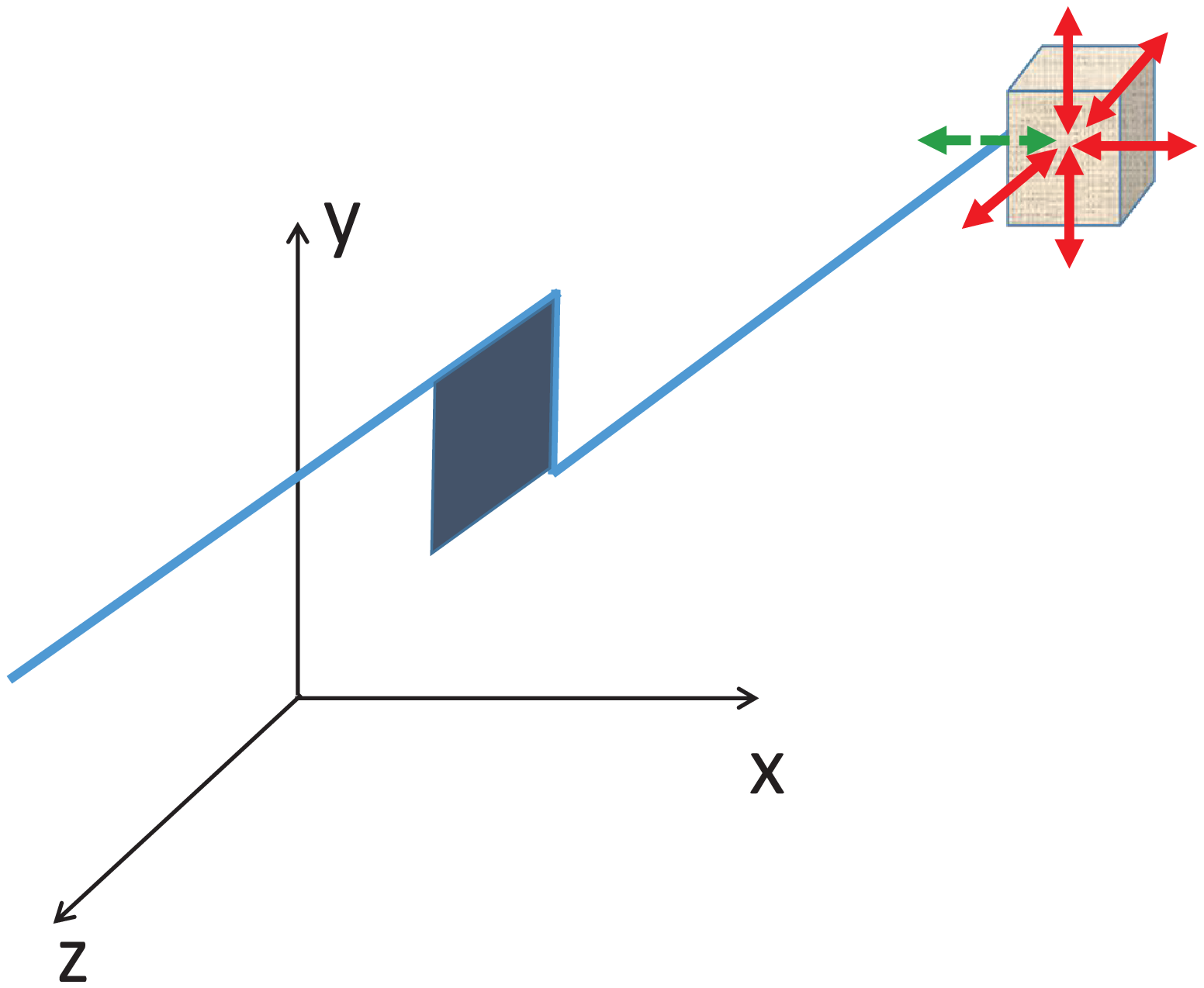} \label{pipe1}}
 \subfigure[]
       {\includegraphics[width=8cm]{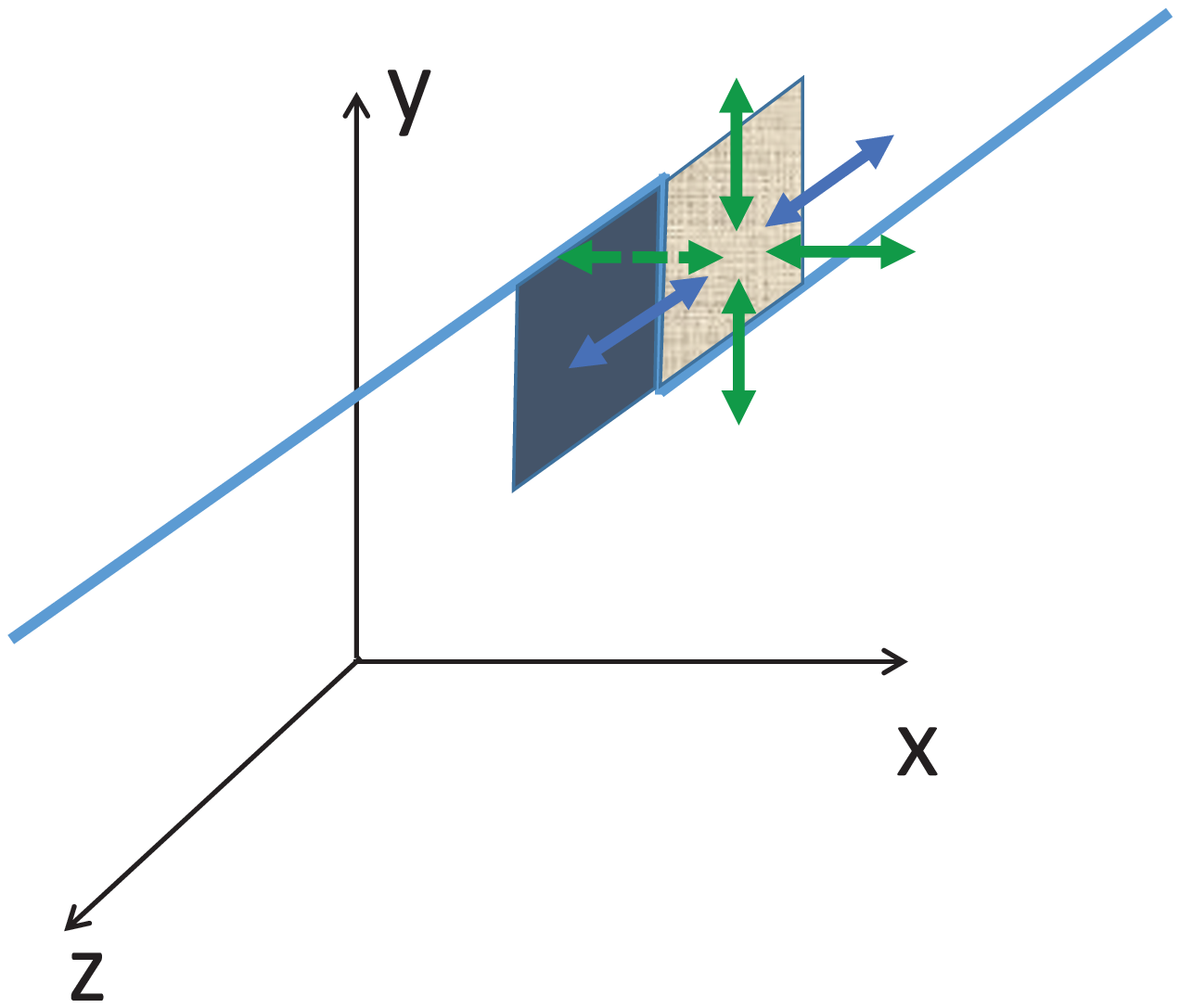} \label{jog1}}
 \subfigure[]
       {\includegraphics[width=8cm]{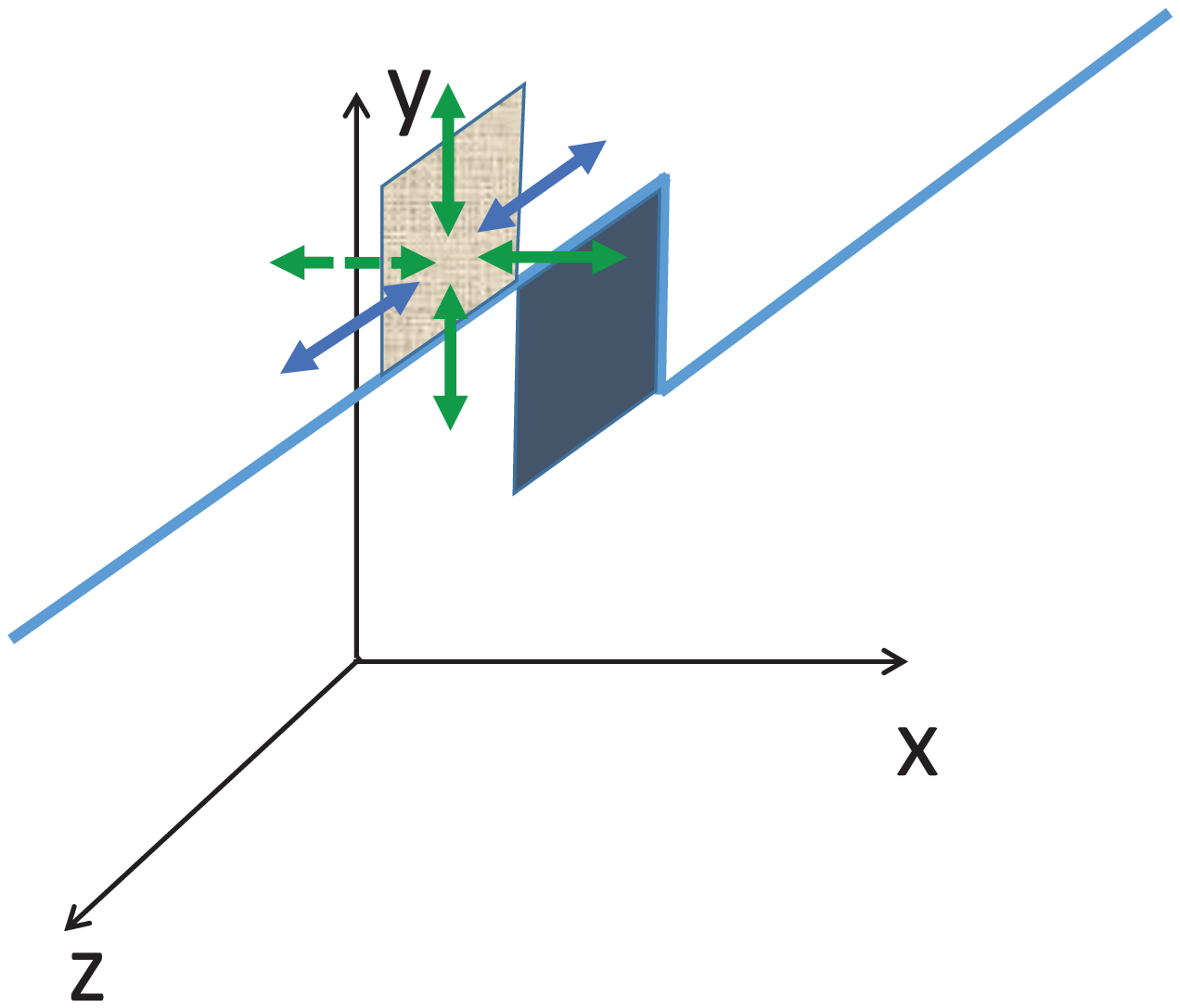} \label{jog2}}
 \subfigure[]
       {\includegraphics[width=8cm]{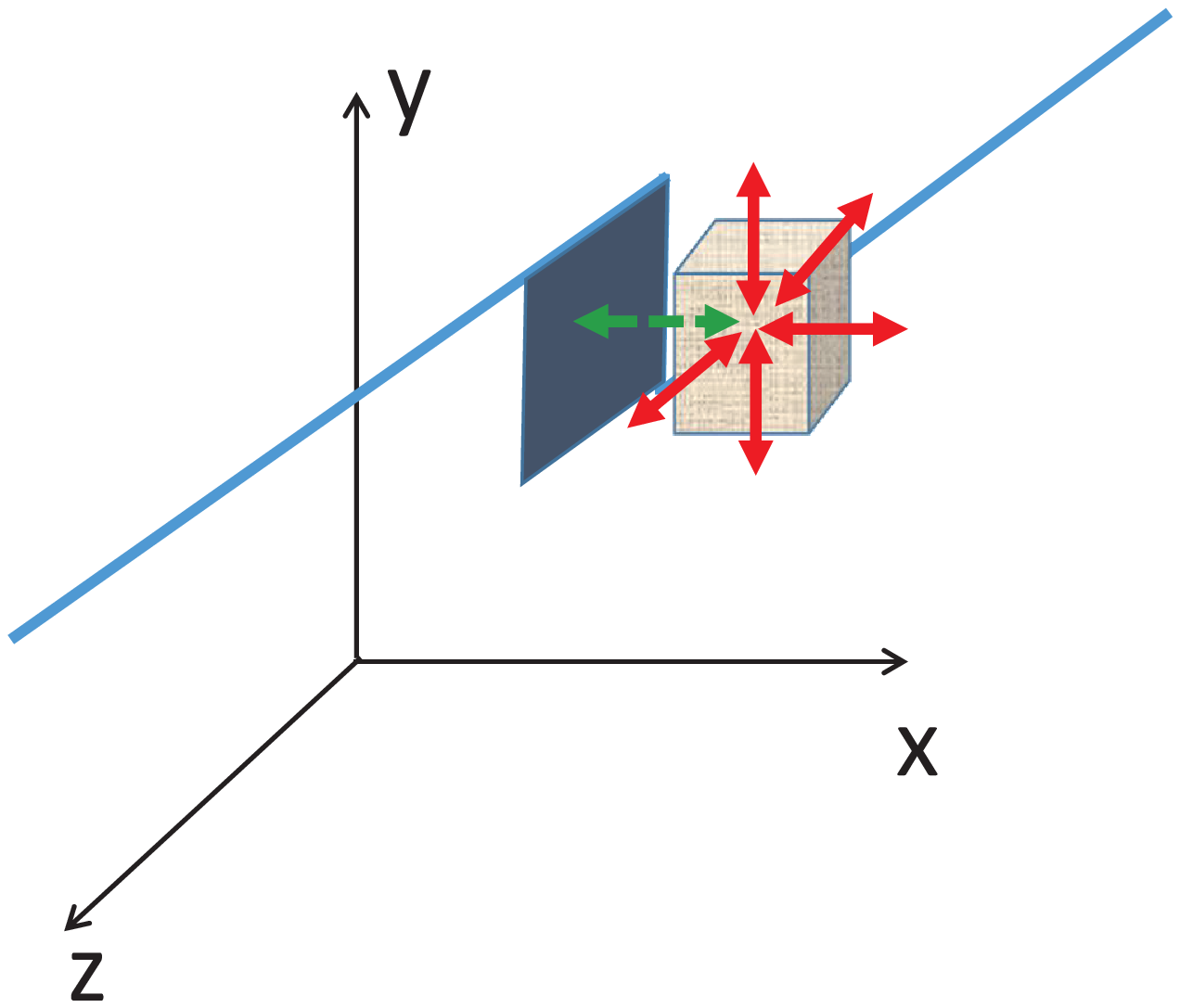} \label{jogout}}
 \subfigure[]
       {\includegraphics[width=8cm]{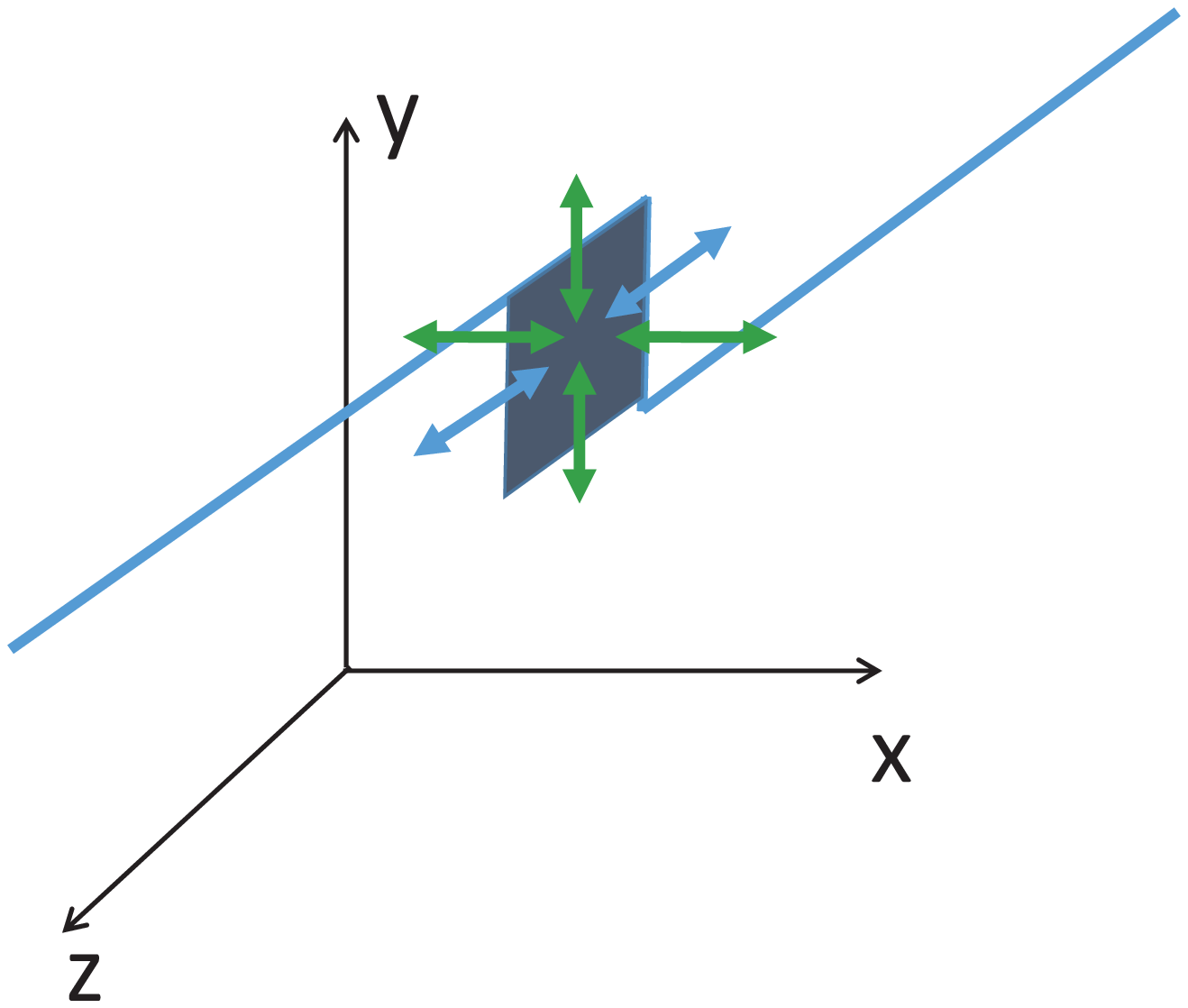} \label{jogout1}}
   \caption{Hopping of vacancies (a) in the bulk away from the dislocation, (b) in the dislocation core away from jogs, (c) adjacent to the dislocation core and away from jogs, (d) and (e) in the dislocation core and adjacent to a jog, (f) in the bulk adjacent to a jog, and (g) at a jog site.}
   \label{fig2}
\end{figure}

\subsubsection{Vacancies in the bulk away from the dislocation}

At a lattice site $(i, j, q)$ sufficiently away from the dislocation, its six nearest neighbors are all bulk sites, and vacancies hop into this site from one of its six neighbors or hop out of this site to one of its six neighbors by the same rate $\Gamma_v$,  see Fig.~\ref{bulk}. These hopping events   follow the standard random walk. From time $t_n$ to $t_{n+1}$, the probability of finding a vacancy at this site  becomes
\begin{eqnarray}\label{eq-bulk-diffusion-discrete}\nonumber
c_{i, j, q}^{v,n+1}&=&(1-6\Gamma_v\tau) c_{i, j, q}^{v,n}+  \Gamma_v\tau( c_{i-1, j, q}^{v,n}+ c_{i+1, j, q}^{v,n}\\
&&+ c_{i, j-1, q}^{v,n}+ c_{i, j+1, q}^{v,n}+c_{i, j, q-1}^{v,n}+c_{i, j, q+1}^{v,n}).
 \end{eqnarray}

\subsubsection{Vacancies in the dislocation core away from jogs}

When a lattice site $(i, j, q)$ is in the dislocation core and sufficiently away from the jogs, it has two nearest neighbors in the dislocation core in the $z$ direction and four nearest neighbors in the bulk in the $x$ and $y$ directions, see Fig.~\ref{pipe}. Following the assumptions in the previous subsection, vacancies may hop from  this site into one of its two neighboring sites in the dislocation core or conversely with rate $\Gamma_c$.  Vacancies may also hop from  this site  into one of the four neighboring sites in the bulk with rate $\Gamma_v \phi_vk_v$,  and in the reverse directions  with rate $\Gamma_v \phi_v$. Thus from time $t_n$ to $t_{n+1}$, the probability of finding a vacancy  at this site  becomes
 \begin{eqnarray}\label{eq-core-diffusion-discrete}\nonumber
 c_{i, j, q}^{c,n+1}&=&(1-2\Gamma_c\tau) c_{i, j, q}^{c,n}+\Gamma_c\tau c_{i, j, q-1}^{c,n}+\Gamma_c\tau c_{i, j, q+1}^{c,n}\\
 &+&\Gamma_v\phi_v\tau( c_{i-1, j, q}^{v,n}+ c_{i+1, j, q}^{v,n}+ c_{i, j-1, q}^{v,n}+ c_{i, j+1, q}^{v,n})-4\Gamma_v\phi_v k_v\tau c_{i, j, q}^{c,n}.
 \end{eqnarray}

 At a neighboring site in the bulk to this site $(i, j, q)$, for example, at the lattice site $(i+1, j, q)$, see Fig.~\ref{pipe1}, we have
 \begin{eqnarray}\label{eq-adjacentcore-diffusion-discrete}\nonumber
 c_{i+1, j, q}^{v,n+1}&=&(1-5\Gamma_v\tau) c_{i+1, j, q}^{v,n}+\Gamma_v\phi_v k_v\tau  c_{i, j, q}^{c,n}-\Gamma_v\phi_v\tau c_{i+1, j, q}^{v,n} \\
 &~&+\Gamma_v\tau( c_{i+2, j, q}^{v,n}+ c_{i+1, j-1, q}^{v,n}+ c_{i+1, j+1, q}^{v,n}+ c_{i+1, j, q-1}^{v,n}+c_{i+1, j, q+1}^{v,n}).
  \end{eqnarray}

\subsubsection{Vacancies in the dislocation core adjacent to a jog}

Now we consider a lattice site $(i, j, q)$ that is in the dislocation core and is adjacent to a jog as shown in Fig.~\ref{jog1} or \ref{jog2}. In this case, vacancies at this site  may hop into or from the jog, or hop into and from another neighboring site in the dislocation core with rate $\Gamma_c$.
Recall that the probability of finding a vacancy at the jog site is $c_J=c_0e^{-\frac{f_{\rm cl}\Omega}{bkT}}$.
 Vacancies may also hop from  this site $(i, j, q)$ into one of the four neighboring sites in the bulk with rate $\Gamma_v \phi_vk_v$,  and in the reverse directions  with rate $\Gamma_v \phi_v$.
Summarizing these events, we have, for example when the jog site is at $(i, j, q+1)$ as in Fig.~\ref{jog1},
 \begin{eqnarray}\label{eq-right-of-jog-diffusion-discrete}\nonumber
 c_{i, j, q}^{c,n+1}&=&(1-2\Gamma_c\tau) c_{i, j, q}^{c,n}+\Gamma_c\tau c_{i, j, q-1}^{c,n}+\Gamma_c\tau c_J\\
 &~&+\Gamma_v\phi_v\tau( c_{i-1, j, q}^{v,n}+ c_{i+1, j, q}^{v,n}+ c_{i, j-1, q}^{v,n}
 + c_{i, j+1, q}^{v,n})-4\Gamma_v\phi_v k_v\tau c_{i, j, q}^{c,n}.
 \end{eqnarray}
The formula for the case when the jog site is at $(i, j, q-1)$ shown in Fig.~\ref{jog2} is similar.

 At an adjacent lattice site in the bulk of this site $(i, j, q)$ (similar to Fig.~\ref{pipe1}), for example, at the site $(i+1, j, q)$,  Eq.~(\ref{eq-adjacentcore-diffusion-discrete}) still holds.

\subsubsection{Vacancies in the bulk adjacent to a jog}

When a lattice site $(i, j, q)$ is in the bulk and adjacent to a jog, the other five neighboring sites of it are in the bulk, see an example in Fig.~\ref{jogout}.   Following the assumptions in the previous subsection, vacancies may  hop from  this site into the adjacent jog with rate $\Gamma_v \phi_vk_v$,  and in the reverse direction  with rate $\Gamma_v \phi_v$. Vacancies may also hop into this site from one of its five neighbors in the bulk or hop out of this site to one of its five neighbors in the bulk by rate $\Gamma_v$. As an example when the adjacent jog is at the site $(i-1, j, q)$, the probability of finding a vacancy  at this site can be described by
 \begin{eqnarray}\label{eq-bulk-near-jog-diffusion-discrete}\nonumber
 c_{i, j, q}^{v,n+1}&=&(1-5\Gamma_v\tau) c_{i, j, q}^{v,n}+\Gamma_v\phi_vk_v\tau c_J-\Gamma_v\phi_v \tau c_{i, j, q}^{v,n} \\
 &~&+\Gamma_v\tau( c_{i+1, j, q}^{v,n}+ c_{i, j-1, q}^{v,n}+ c_{i, j+1, q}^{v,n}+ c_{i, j, q-1}^{v,n}+ c_{i, j, q+1}^{v,n}).
 \end{eqnarray}

 \subsubsection{Stochastic motion of a jog}

 \begin{figure}[htbp]
  \centering
    \subfigure[]
       {\includegraphics[width=7cm]{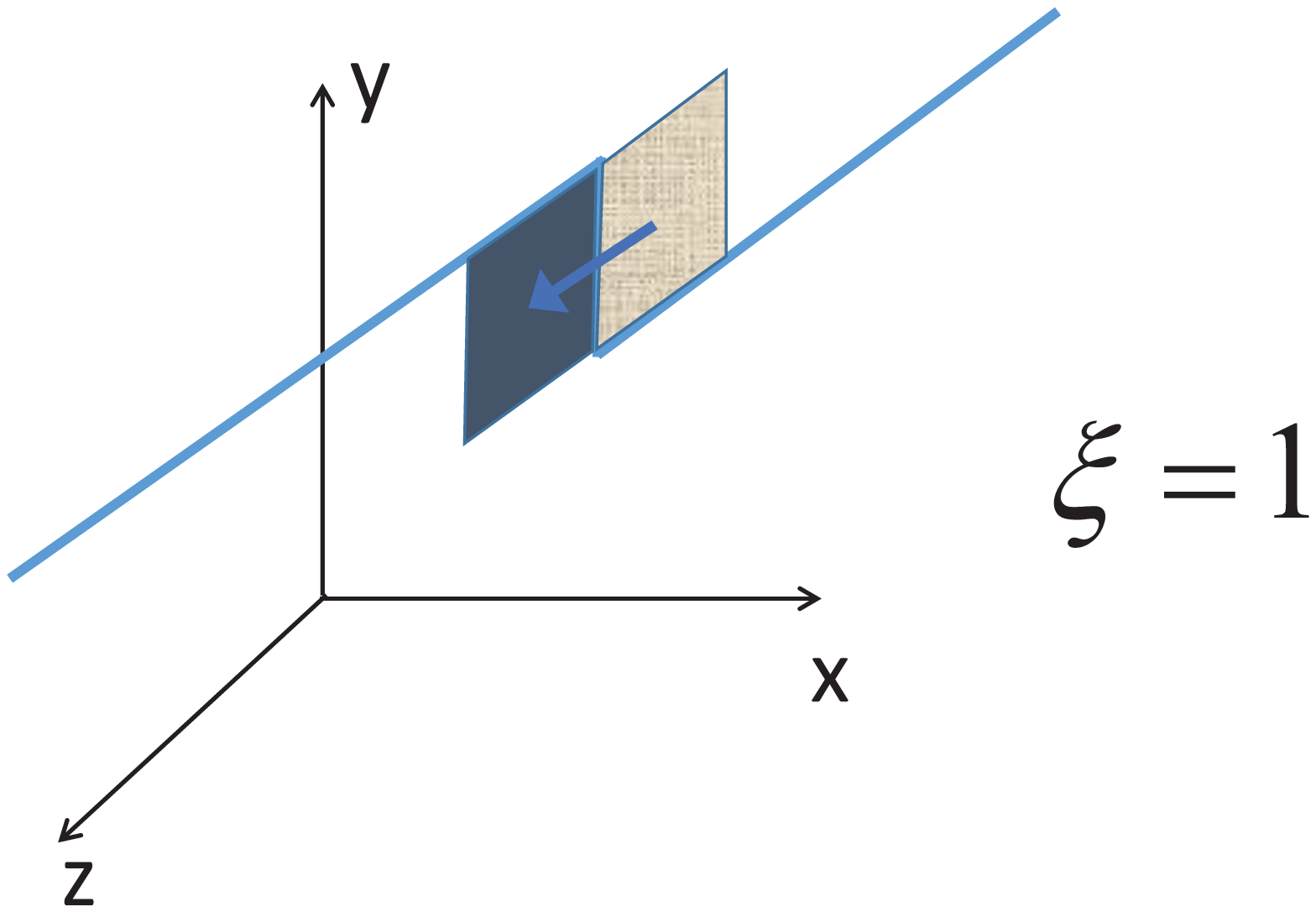} \label{stochastic21}}
  \subfigure[]
       {\includegraphics[width=7cm]{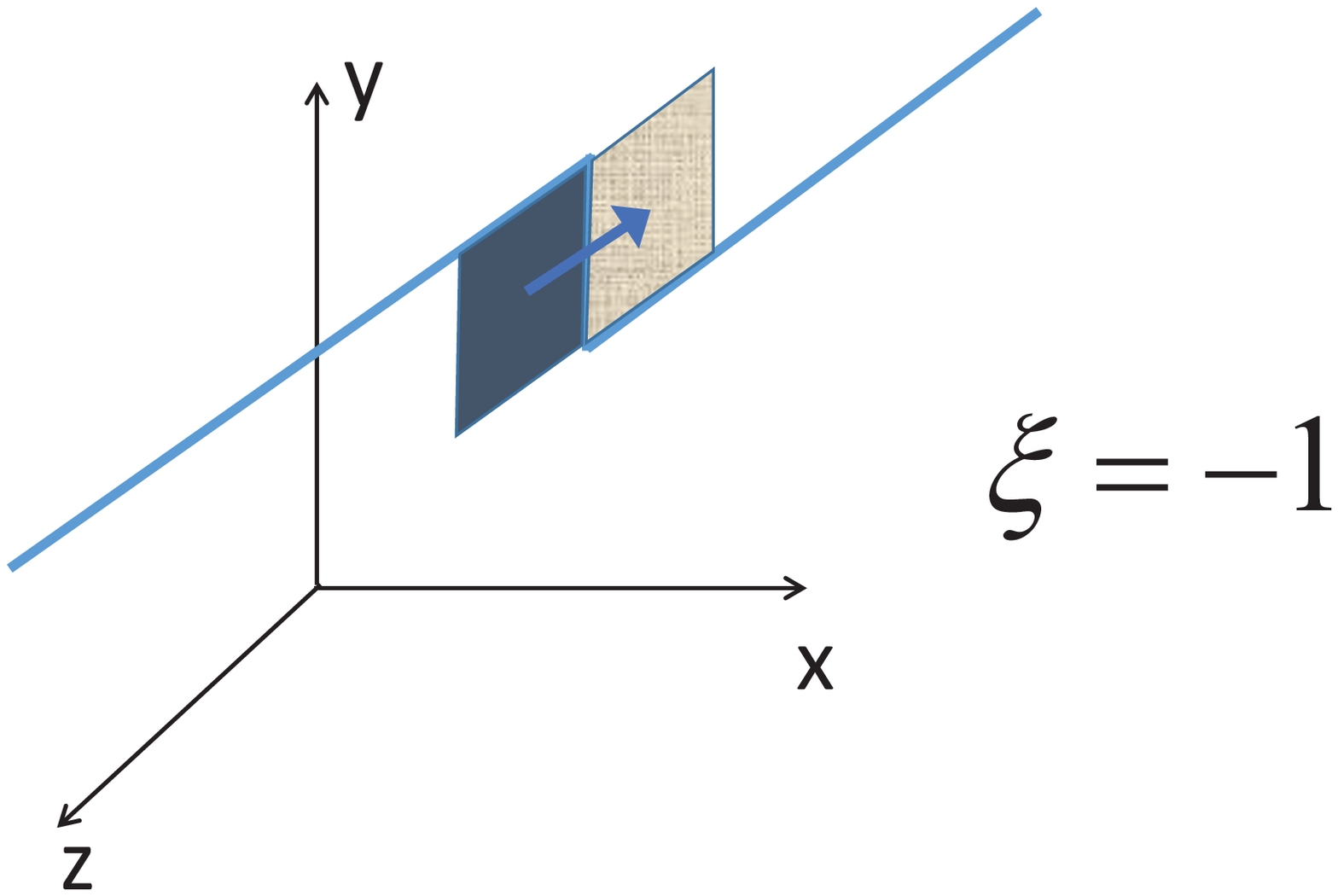} \label{stochastic22}}
  \subfigure[]
       {\includegraphics[width=7cm]{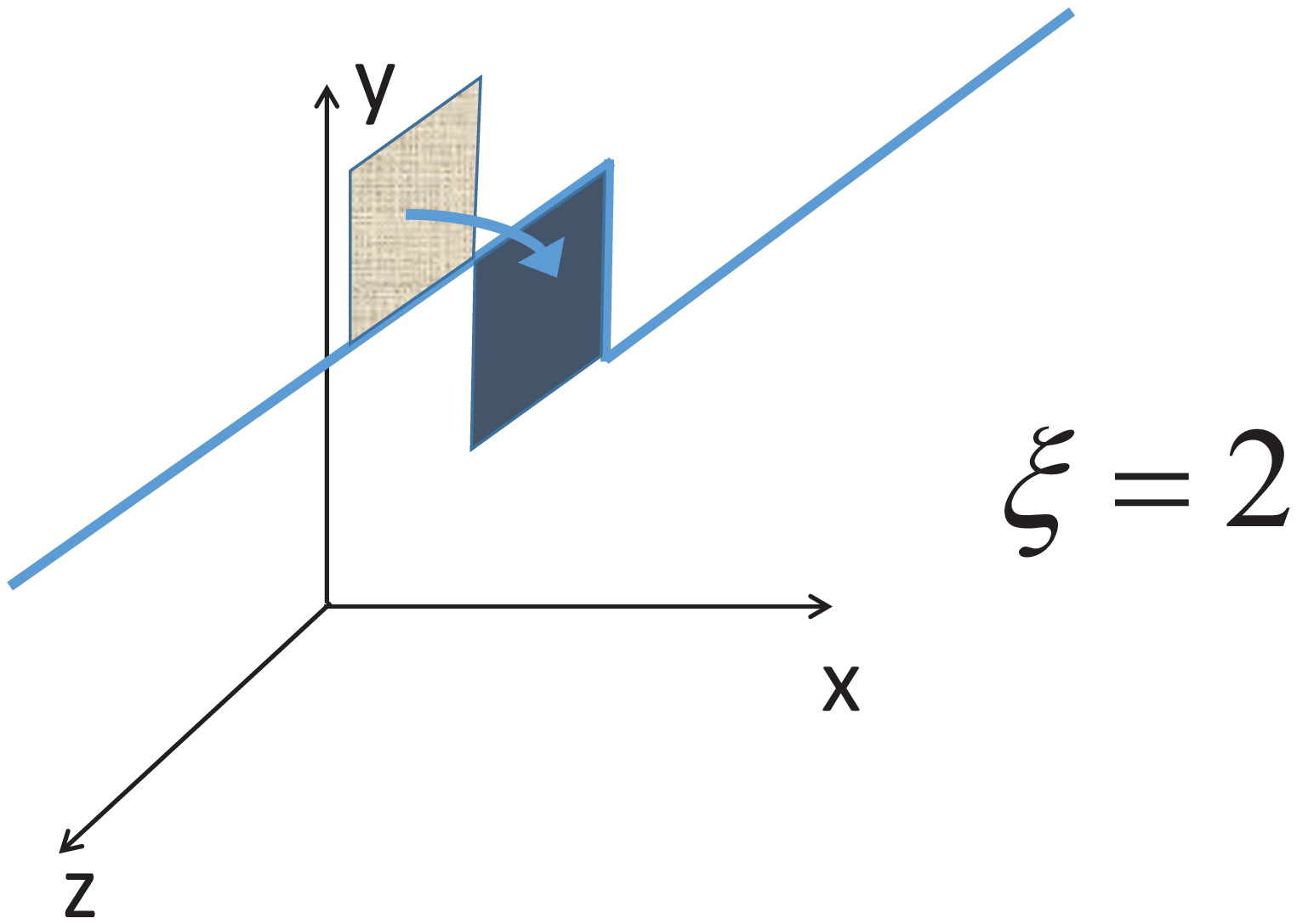} \label{stochastic31}}
   \subfigure[]
       {\includegraphics[width=7cm]{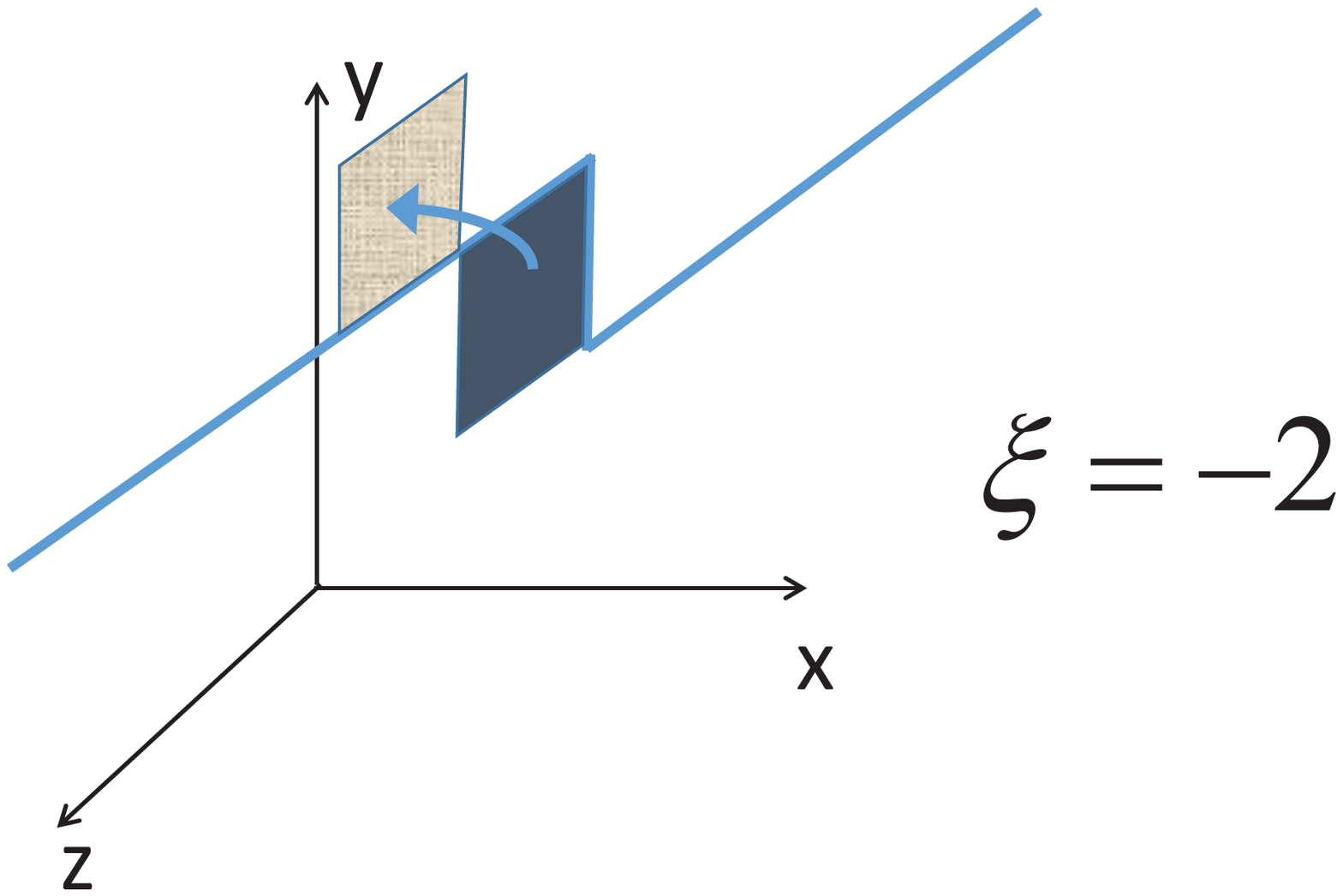} \label{stochastic32}}
 \subfigure[]
       {\includegraphics[width=7cm]{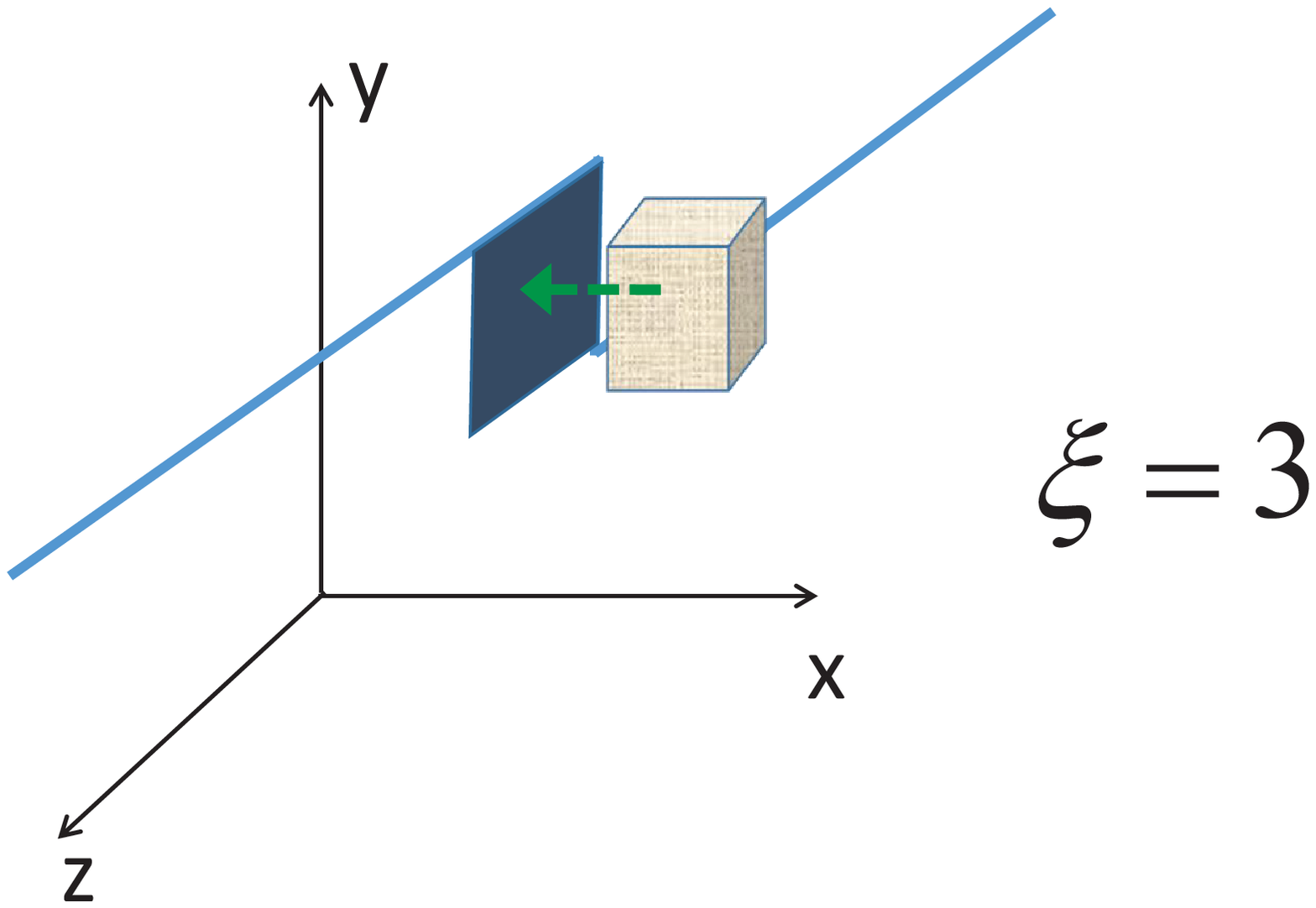} \label{stochastic41}}
\subfigure[]
       {\includegraphics[width=7cm]{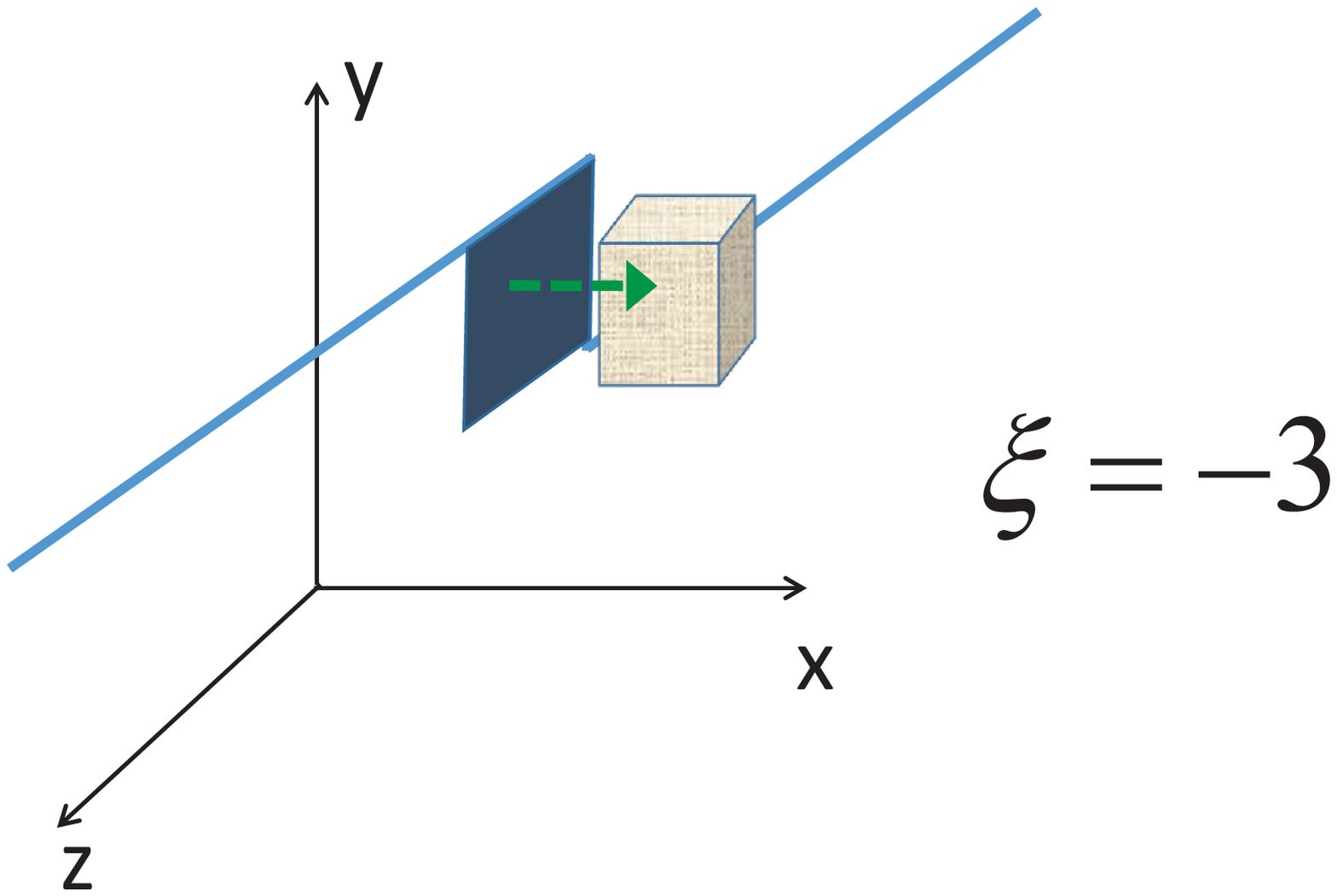} \label{stochastic42}}
    \subfigure[]
       {\includegraphics[width=7cm]{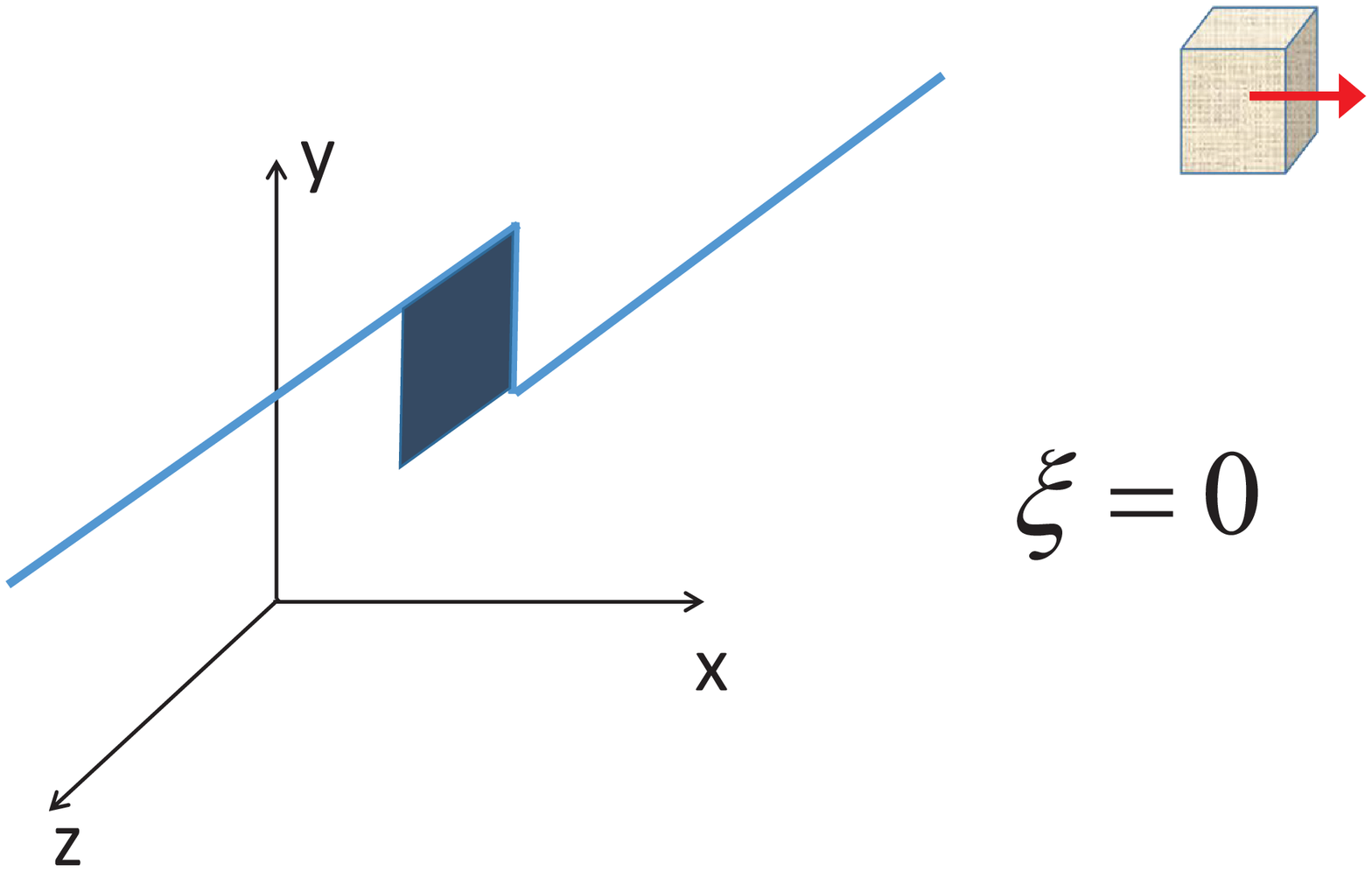}\label{stochastic1}}
  \caption{Motion of a jog determined by a random variable $\xi$ at each time step $t_n$. (a) $\xi=1$ and (c) $\xi=2$:  the jog moves forward by one lattice site due to vacancies hopping from a neighboring site in the dislocation core. (b) $\xi=-1$ and (d) $\xi=-2$:  the jog moves backward by one lattice site by emitting vacancies to a neighboring site in the dislocation core. (e) $\xi=3$: the jog moves forward by one lattice site due to vacancies hopping from the $+x$-neighboring site  in the bulk. (f) $\xi=-3$: the jog moves  backward by one lattice site by emitting vacancies to the $+x$ neighboring site on the $+x$ side to the dislocation in the bulk. The cases of $\xi=\pm 4$, $\xi=\pm 5$, and $\xi=\pm 6$ are defined similarly due to exchange of vacancies with the $-x$-neighboring, $+y$-neighboring, and $-y$-neighboring sites, respectively. (g) $\xi=0$: the jog does not move.
  }
   \label{fig_sto}
\end{figure}

We consider the stochastic motion of jogs.
The motion of the jog $z_m$ is determined by a random variable $\xi_m$ associated with it. At the discrete time $t_n$, when the jog $z_m$ is located at the site $(i,j,q)$, the random variable $\xi_m$ takes its value from the set ${\mathcal S}=\{0,\pm 1,\pm 2, \pm 3, \pm 4, \pm 5, \pm 6\}$ according to the following probabilities
\begin{equation}
\begin{array}{ll}
{\mathbb P}(\xi_m(t_n)=1)=\Gamma_cc_{i, j, q-1}^{c,n}\tau, & {\mathbb P}(\xi_m(t_n)=-1)=\Gamma_cc_J\tau,\\
{\mathbb P}(\xi_m(t_n)=2)=\Gamma_cc_{i, j, q+1}^{c,n}\tau, & {\mathbb P}(\xi_m(t_n)=-2)=\Gamma_cc_J\tau,\\
{\mathbb P}(\xi_m(t_n)=3)=\Gamma_v\phi_vc_{i+1, j, q}^{v,n}\tau, & {\mathbb P}(\xi_m(t_n)=-3)=\Gamma_v\phi_vk_vc_J\tau,\\
{\mathbb P}(\xi_m(t_n)=4)=\Gamma_v\phi_vc_{i-1, j, q}^{v,n}\tau, & {\mathbb P}(\xi_m(t_n)=-4)=\Gamma_v\phi_vk_vc_J\tau,\\
{\mathbb P}(\xi_m(t_n)=5)=\Gamma_v\phi_vc_{i, j+1, q}^{v,n}\tau, & {\mathbb P}(\xi_m(t_n)=-5)=\Gamma_v\phi_vk_vc_J\tau,\\
{\mathbb P}(\xi_m(t_n)=6)=\Gamma_v\phi_vc_{i, j-1, q}^{v,n}\tau, & {\mathbb P}(\xi_m(t_n)=-6)=\Gamma_v\phi_vk_vc_J\tau,\\
{\mathbb P}(\xi_m(t_n)=0)=1-{\displaystyle \sum_{p\in {\mathcal S}, p\neq 0}} {\mathbb P}(\xi_m(t_n)=p). &
\end{array}
\label{eqn:jobprob}
\end{equation}

The events corresponding to these values of $\xi_m(t_n)$ are shown schematically in Fig.~\ref{fig_sto}. Now we explain these events and the probabilities in Eq.~\eqref{eqn:jobprob}. When $\xi_m(t_n)=1$, the jog $z_m$ located at the site $(i,j,q)$ moves by one lattice site to $(i,j,q-1)$ by absorbing vacancies hopping from the  neighboring site  $(i,j,q-1)$ in the dislocation core, see Fig.~\ref{stochastic21}. Following the assumptions  in the previous subsection, the rate of such hopping for a vacancy is $\Gamma_c$. Thus the probability of this event is $\Gamma_cc_{i, j, q-1}^{c,n}\tau$. When $\xi_m(t_n)=-1$, this jog  moves by one lattice site to $(i,j,q+1)$ by emitting vacancies into the  neighboring site  $(i,j,q-1)$ in the dislocation core, see Fig.~\ref{stochastic22}. The rate of such hopping for a vacancy is also $\Gamma_c$. Using the probability of finding a vacancy  $c_J=c_0e^{-\frac{f_{\rm cl}\Omega}{bkT}}$ at the jog site, the probability of the event $\xi_m(t_n)=-1$ is $\Gamma_cc_J\tau$. Similarly, the events of $\xi_m(t_n)=2$ and $-2$ and their probabilities are used to describe the motion of the jog due to exchange of vacancies with the neighboring site $(i,j,q+1)$, as shown in Fig.~\ref{fig_sto}(c) and (d).

 When $\xi_m(t_n)=3$, the jog  moves by one lattice site to $(i,j,q-1)$ by absorbing vacancies hopping from the  neighboring site  $(i+1,j,q)$ in the bulk, see Fig.~\ref{stochastic41}. Recalling that the rate of such hopping for a vacancy is $\Gamma_v\phi_v$, the probability of this event is $\Gamma_v\phi_vc_{i+1, j, q}^{v,n}\tau$. Fig.~\ref{stochastic42} shows the case of  $\xi_m(t_n)=-3$, in which the jog moves  by one lattice site to $(i,j,q+1)$ by emitting vacancies into the  neighboring site  $(i+1,j,q)$ in the bulk. This event, following the hopping rate $\Gamma_v\phi_vk_v$ for a vacancy, has the probability $\Gamma_v\phi_vk_vc_J\tau$, where again we have used the probability of finding a vacancy  $c_J$ at the jog site. The events of $\xi_m(t_n)=\pm 4$, $\xi_m(t_n)=\pm 5$, and $\xi_m(t_n)=\pm 6$ are defined and their probabilities are calculated similarly, due to exchange of vacancies with the $-x$-neighboring, $+y$-neighboring, and $-y$-neighboring sites, respectively. Finally, $\xi_m(t_n)=0$ summarizes the rest cases where the jog does not move, see Fig.~\ref{stochastic1}.

Whenever the jog moves, the probability of finding a vacancy  at the new site is set to be the equilibrium value $c_J$, and the probability of finding a vacancy  at the original site of the jog remains to be $c_J$.

The motion of the jog $z_m$ from $t_n$ to $t_{n+1}$ is then, for an upward jog as shown in Fig.~\ref{fig_sto}, given as follows.
\begin{equation}
z_m(t_{n+1})=\left\{
\begin{array}{ll}
z_m(t_n)-b &{\rm if}\ \xi>0,\\
z_m(t_n)+b &{\rm if}\ \xi<0,\\
z_m(t_n) &{\rm if}\ \xi=0.
\end{array}
\right.
\end{equation}
For a downward jog, the motion is in the opposite direction with the same events.

We can then calculate the mean displacement of the jog  within time $\tau=t_{n+1}-t_n$ as
 \begin{eqnarray}\label{mean-value-jog-motion-discrete}\nonumber
 E(z_m(t^{n+1})-z_m(t^n)) &=& -\tau b\big[\Gamma_c (c_{i, j, q+1}^{c,n}+ c_{i, j, q-1}^{c,n})
 -\Gamma_v\phi_v( c_{i-1, j, q}^{v,n}+ c_{i+1, j, q}^{v,n}+ c_{i, j-1, q}^{v,n}+ c_{i, j+1, q}^{v,n})\\
 &&+(2\Gamma_c+4\Gamma_v\phi_vk_v)c_J\big].
  \end{eqnarray}
  Thus the velocity of the jog in its motion direction (the $-z$ direction for an upward jog or the $+z$ direction for a downward jog) is
\begin{eqnarray}\label{velocity-jog-motion-discrete}\nonumber
 V_{\rm jog} &=& \Gamma_c b(c_{i, j, q+1}^{c,n}+ c_{i, j, q-1}^{c,n})
 +\Gamma_vb\phi_v( c_{i-1, j, q}^{v,n}+ c_{i+1, j, q}^{v,n}+ c_{i, j-1, q}^{v,n}+ c_{i, j+1, q}^{v,n})\\
 &&-(2\Gamma_cb+4\Gamma_vb\phi_vk_v)c_J.
  \end{eqnarray}

  In fact, by conservation of vacancies, i.e. the vacancies that are absorbed or emitted by the jog for its motion equals the vacancies that are lost at the rest non-jog sites within the time interval $[t_n,t_{n+1}]$, using our scheme, the jog velocity should satisfy
\begin{eqnarray}\nonumber
 \hat{V}_{\rm jog} &=& \Gamma_c b(c_{i, j, q+1}^{c,n}+ c_{i, j, q-1}^{c,n})\cdot[1+(c_{i, j, q-1}^{c,n}-c_J)]\\ \nonumber
&& +\Gamma_vb\phi_v( c_{i-1, j, q}^{v,n}+ c_{i+1, j, q}^{v,n}+ c_{i, j-1, q}^{v,n}+ c_{i, j+1, q}^{v,n})\cdot[1+(c_{i, j, q-1}^{c,n}-c_J)]\\
 &&-(2\Gamma_cb+4\Gamma_vb\phi_vk_v)c_J\cdot[1+(c_{i, j, q+1}^{c,n}-c_J)].
  \end{eqnarray}
 The differences  $c_{i, j, q-1}^{c,n}-c_J$ and  $c_{i, j, q+1}^{c,n}-c_J$ are due to the treatment of resetting the probability of finding a vacancy  at the new jog site from its original value $c_{i, j, q-1}^{c,n}$ or $c_{i, j, q+1}^{c,n}$ to the new value $c_J$. These differences are small since these probabilities are small compared with $1$. Thus the jog velocity formula in Eq.~\eqref{velocity-jog-motion-discrete} provides a good approximation to this exact expression.

\section{The jog dynamics model}\label{sec:III}

The jog dynamics model for dislocation climb is obtained by letting $b\rightarrow 0$ and $\tau\rightarrow 0$ in the microscopic model presented in the previous section.

 \begin{figure}[htbp]
 \centering
 \includegraphics[width=15cm]{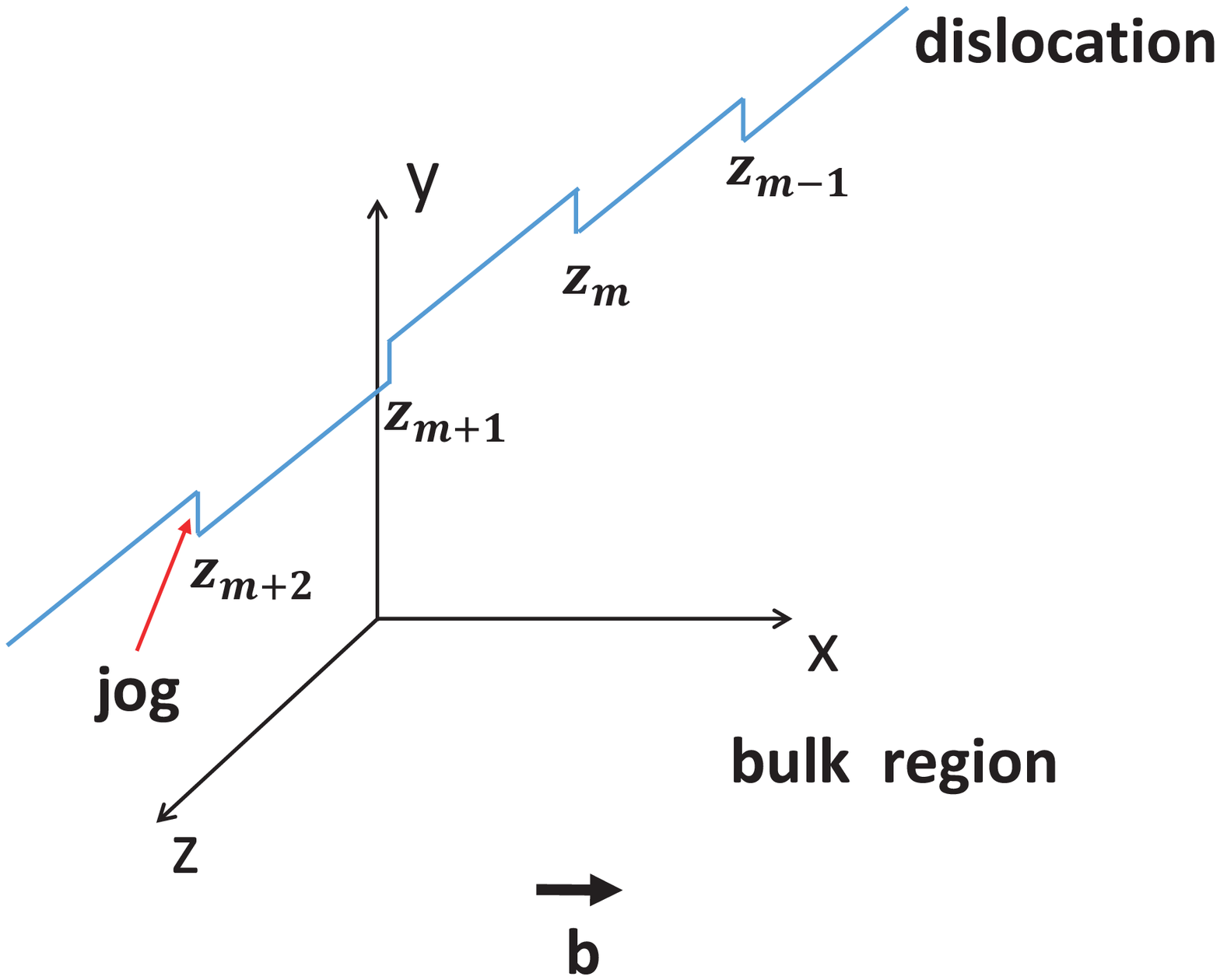}
  \caption{The jog dynamics model for dislocation climb.}
  \label{fig:jogdynamics}
\end{figure}

This model is based upon continuous spatial and time domains $(x,y,z)\in \mathbf R^3$ and $t\in [0,\infty)$. The dislocation is a jogged straight line nominally parallel to the $z$ axis, see Fig.~\ref{fig:jogdynamics}. The dislocation has  core radius $r_d$  and Burgers vector $\mathbf b$ in the $+x$ direction. The region outside the dislocation core is the bulk region.
 The vacancy concentrations in the bulk and in the dislocation core are denoted respectively by $c^v(x,y,z,t)$ if $(x,y,z)$ is in the bulk at time $t$, and $c^c(z,t)$ if the point is within the dislocation core at time $t$. When $(x,y,z)=(ib, jb, qb)$ and $t=n\tau$, the vacancy concentrations $c^v$ and $c^c$ are related to those in the microscopic model in the previous section by  $c^v(x,y,z,t)=c_{i, j, q}^{v,n}$ and  $c^c(z,t)=c_{i, j, q}^{c,n}$, respectively. Note that we assume that the vacancy concentration within the dislocation core  is uniform in the cross-section of the dislocation core, thus in space $c^c$ depends only on $z$ which is a parametrization of the dislocation line.

After taking Taylor expansions at the point $(i, j, q)$ in the bulk and time $t_n$ in terms of the small parameters $b$ and $\tau$, and keeping the leading order contributions, Eq.~(\ref{eq-bulk-diffusion-discrete}) becomes the following diffusion equation in the bulk:
 \begin{eqnarray}\label{eq-bulk-diffusion-continuum}
c_t^v=D_v \big(c_{xx}^v+c_{yy}^v+c_{zz}^v\big),
 \end{eqnarray}
where the bulk diffusion constant $D_v=\Gamma_v b^2$, { and $c^v_t={\partial c^v}/{\partial t}$, $c_{xx}^v={\partial^2 c^v}/{\partial x^2}$ are partial derivatives (similar notations for other partial derivatives).}

Similarly in this continuum limit,
 Eq.~(\ref{eq-core-diffusion-discrete}) becomes the following one-dimensional diffusion equation along the dislocation, which is the pipe diffusion for vacancies:
 \begin{eqnarray}\label{eq-core-diffusion-continuum}
c_t^c=D_c c_{zz}^c+\frac{4D_v\phi_v}{b^2}(c^v-k_vc^c),
 \end{eqnarray}
where the pipe diffusion constant $D_c=\Gamma_c b^2$.

Next, we consider the continuum limit of Eq.~(\ref{eq-adjacentcore-diffusion-discrete}) in the microscopic model in the previous section, which holds at a neighboring site in the $+x$ direction in the bulk to the dislocation core, see Fig.~\ref{pipe1}.
 For convenience of obtaining the continuum limit, we make an extension of the vacancy concentration in the bulk by bulk diffusion to the dislocation core  and get a virtual vacancy concentration $c_{i, j, q}^{v,n}$ at the dislocation core site. Eq.~(\ref{eq-adjacentcore-diffusion-discrete}) then becomes
   \begin{eqnarray}\label{eq-adjacentcore-diffusion-discrete-1}\nonumber
 c_{i+1, j, q}^{v,n+1}&=&(1-6\Gamma_v\tau) c_{i+1, j, q}^{v,n}\\ \nonumber
 &~&+\Gamma_v\tau( c_{i, j, q}^{v,n}+c_{i+2, j, q}^{v,n}+ c_{i+1, j-1, q}^{v,n}+ c_{i+1, j+1, q}^{v,n}+ c_{i+1, j, q-1}^{v,n}+c_{i+1, j, q+1}^{v,n})\\
 &~& +\Gamma_v\tau(c_{i+1, j, q}^{v,n}- c_{i, j, q}^{v,n})-\Gamma_v\phi_v\tau (c_{i+1, j, q}^{v,n}- k_v  c_{i, j, q}^{c,n}).
  \end{eqnarray}
  Now the first two lines in this equation takes the same form as the bulk diffusion scheme in Eq.~(\ref{eq-bulk-diffusion-discrete}) at the site $(i+1, j, q)$, thus its continuum limit is the bulk diffusion equation as we have obtained above in Eq.~(\ref{eq-bulk-diffusion-continuum}). Keeping the leading order contributions in the continuum limit for the rest two terms in the last line in Eq.~(\ref{eq-adjacentcore-diffusion-discrete-1}), we get the continuum limit of
  Eq.~(\ref{eq-adjacentcore-diffusion-discrete-1}):
 \begin{eqnarray}\label{eq-adjacentcore-diffusion-continuum}
c_t^v&=&D_v(c_{xx}^v+c_{yy}^v+c_{zz}^v)+\Gamma_v b c_x^v -\Gamma_v\phi_v(c^v-k_vc^c).
 \end{eqnarray}
  Since the bulk vacancy concentration $c^v$ satisfies the  diffusion equation (\ref{eq-bulk-diffusion-continuum}), Eq.~(\ref{eq-adjacentcore-diffusion-continuum}) results in the following equation which in fact serves as a boundary condition for the vacancy bulk diffusion when the point approach the dislocation core surface from the $+x$ direction:
 \begin{eqnarray}\label{eq-boundary-bulk-continuum}
D_v c_x^v =\frac{D_v\phi_v}{b}\big(c^v-k_vc^c\big).
 \end{eqnarray}

When the site is adjacent to a jog in the $+x$ direction in the bulk as shown in Fig.~\ref{jogout}, using the above treatment, we obtain the following continuum limit of Eq.~(\ref{eq-bulk-near-jog-diffusion-discrete}):
   \begin{eqnarray}\label{eq-bulk-near-jog-diffusion-continuum}
c_t^v&=&D_v(c_{xx}^v+ c_{yy}^v+ c_{zz}^v)+\Gamma_v b c_x^v-\Gamma_v\phi_v(c^v-k_vc_J).
 \end{eqnarray}
Again, using the bulk diffusion equation (\ref{eq-bulk-diffusion-continuum}), we have  the boundary condition for the vacancy bulk diffusion near a jog in the $+x$ direction:
  \begin{eqnarray}\label{eq-boundary-bulk-near-jog-continuum}
D_v c_x^v =\frac{D_v\phi_v}{b}\big(c^v-k_vc_J\big).
 \end{eqnarray}

 For a site in the dislocation core and adjacent to a jog in the $-z$ side as shown in Fig.~\ref{jog1}, whose microscopic scheme is given by Eq.~(\ref{eq-right-of-jog-diffusion-discrete}), its continuum limit is also Eq.~(\ref{eq-core-diffusion-continuum}). Here we have used the equilibrium vacancy concentration at the jog site $(i,j,q+1)$ that $c^{c,n}_{i,j,q+1}=c_J=c_0^ce^{-\frac{f_{\rm cl}\Omega}{bkT}}$, which in the continuum limit model is the boundary condition
\begin{equation}
c^c=c_J=c_0^ce^{-\frac{f_{\rm cl}\Omega}{bkT}},
\end{equation}
at a jog location $(x,y,z)$ and any time $t$. These arguments also hold when the site in the dislocation core is next to a jog site in the $+z$ direction as shown in Fig.~\ref{jog2}.

Finally, from Eq.~(\ref{velocity-jog-motion-discrete}), the continuum limit of the velocity of the jog located at $z_m$ is
  \begin{eqnarray}\label{velocity-jog1}
  v_{\rm jog}^{(m)} =D_c\big[c^c_z(z_m^+)-c^c_z(z_m^-)\big]+\left.\frac{4D_v\phi_v}{b}\big(c^v-k_vc_J\big)\right|_{z=z_m}.
  \end{eqnarray}

{ Note that the above formulas
 are derived from the microscopic scheme presented in the previous section, which is based on the simple cubic lattice with dislocation core width being the length of the Burgers vector. Real crystalline materials, e.g. metals, commonly have  fcc, bcc or hcp  lattices, and dislocations in  different materials have different core sizes.
 The above derived jog dynamics model can be applied beyond the simple cubic lattice if we generalize the formulation using dislocation core radius $r_d$ as follows. The influence of the lattice structures can be incorporated in the parameters in the model.

 In this general case, the bulk diffusion vacancy flux into the dislocation is
  \begin{equation}
  \mathbf j\cdot \mathbf n|_{r=r_d}=-D_v\nabla c^v\cdot\mathbf n|_{r=r_d}=-\left.D_v \frac{\partial c^v}{\partial n}\right|_{r=r_d},
   \end{equation}
   where $\mathbf n$ is the inward normal direction on the dislocation core surface $r=r_d$, and $r$ is the distance to the dislocation. The pipe diffusion vacancy flux into the jog is
   \begin{equation}
   \mathbf j^c\cdot \mathbf n^c|_{z=z_m^+, z_m^-}=-D_c c^c_{z}(\mathbf n^c\cdot \hat{\mathbf z})|_{z=z_m^+, z_m^-},
    \end{equation}
    where  $\mathbf n^c$ is the direction into the jog which is in the $-z$ direction from the positive side of the jog on the dislocation and in the $+z$ direction from the negative side, and  $\hat{\mathbf z}$ is the unit vector in the $+z$ direction.

     Recall that  Eq.~(\ref{eq-boundary-bulk-continuum}) is for the case when the point approaches dislocation core from the $+x$ axis. In the general case at a point on the surface of the dislocation core $r=r_d$,  the condition is
    \begin{equation}
    {\displaystyle -D_v \frac{\partial c^v}{\partial n} =\left.\frac{D_v\phi_v}{b}\big(c^v-k_vc^c\big)\right|_{r=r_d}}.
    \end{equation}
Also recall  that the last term  in Eq.~(\ref{eq-core-diffusion-continuum}) (or (\ref{velocity-jog1})) is the total bulk diffusion flux into the dislocation core (or a jog) based on the simple cubic lattice. When the dislocation core surface is $r=r_d$ instead of the four discrete sites in the simple cubic lattice, this term describing the total bulk diffusion flux into the dislocation core  should be
$\frac{1}{b^2}\int_{r=r_d} \mathbf j\cdot \mathbf n \ dl=-\frac{1}{b^2}\int_{r=r_d} D_v \frac{\partial c^v}{\partial n}dl$ in Eq.~(\ref{eq-core-diffusion-continuum}), or
$\frac{1}{b}\int_{r=r_d} \mathbf j\cdot \mathbf n \ dl=-\frac{1}{b}\int_{r=r_d} D_v \frac{\partial c^v}{\partial n}dl$ at the jog $z_m$
in Eq.~(\ref{velocity-jog1}).}

  Summarizing these equations with the above generalizations beyond the simple cubic lattice, we obtain a vacancy pipe diffusion equation with Dirichlet boundary condition at the jogs  and a vacancy bulk diffusion equation with Robin boundary condition near the dislocation as follows:
\begin{eqnarray}\label{eq-core-diffusion-equation}
    \left\{
        \begin{array}{l}
            c_t^c= D_c c_{zz}^c+\frac{2\pi r_dD_v }{ b^2l_\phi}
            \left(\frac{1}{2\pi r_d}\int_{r=r_d} c^v \ dl-k_vc^c\right), \ {\rm on \ the \ dislocation}, \vspace{1ex}\\
           c^c=\left.c_0^ce^{-\frac{f_{\rm cl}\Omega}{bkT}}\right|_{z=z_m},
        \end{array}
    \right.
\end{eqnarray}
 and
\begin{eqnarray}\label{eq-diffusion-equation}
    \left\{
        \begin{array}{l}
     c_t^v= D_v \big( c_{xx}^v+ c_{yy}^v+ c_{zz}^v\big), \ {\rm in \ the \ bulk},\vspace{1ex}\\
          {\displaystyle -\frac{\partial c^v}{\partial n} =\left.\frac{1}{l_\phi}\big(c^v-k_vc^c\big)\right|_{r=r_d}},  \vspace{1ex}\\
          c^v=\left. c_{\infty} \right|_{r=r_\infty},
        \end{array}
    \right.
\end{eqnarray}
where
\begin{equation}\label{eqn:l_phi_v}
l_\phi=b/\phi_v,
\end{equation}
and $r_\infty$ is the outer cutoff for the distance to the dislocation.
When the solution of bulk diffusion $c^v$ is symmetric with respect to $r$ (e.g. for a straight edge dislocation), the pipe diffusion equation in Eq.~(\ref{eq-core-diffusion-equation}) is reduced to
$c_t^c= D_c c_{zz}^c+\frac{2\pi r_dD_v}{ b^2l_\phi}\left( c^v -k_vc^c\right)$.

The boundary value problem in Eq.~\eqref{eq-core-diffusion-equation} governs the vacancy pipe diffusion  in the dislocation core. The source term in the equation comes from the net vacancy flux into the dislocation core from the bulk. The Dirichlet boundary condition at each jog describes the thermodynamic equilibrium condition there. The boundary value problem in Eq.~\eqref{eq-diffusion-equation} governs the vacancy diffusion in the bulk. The boundary conditions of it include a Robin boundary condition that describes the vacancy flux from the bulk into the dislocation core, and a condition in the far field. The parameter $l_\phi$  is a characteristic length that indicates the barrier for the vacancies hopping into the dislocation core from the bulk with respect to the hopping within the bulk. More physical meanings of $l_\phi$ will be discussed in the dislocation dynamics model in the next section.

From Eq.~(\ref{velocity-jog1}), the continuum limit of the jog velocity in its motion direction in the general setting is
  \begin{eqnarray}\label{velocity-jog2}
  v_{\rm jog}^{(m)}
  &=&\sum_{s=+,-}\mathbf j^c\cdot \mathbf n^c|_{z=z_m^{(s)}}+\left.\frac{1}{b}\int_{r=r_d}\mathbf j\cdot \mathbf n \ dl\right|_{z=z_m} \vspace{1ex}\\
  &=&D_c\big[c^c_z(z_m^+)-c^c_z(z_m^-)\big]+\left.\frac{2\pi r_dD_v}{ bl_\phi}\left(\frac{1}{2\pi r_d}\int_{r=r_d} c^v \ dl-k_v c_0^ce^{-\frac{f_{\rm cl}\Omega}{bkT}}\right)\right|_{z=z_m}
  \end{eqnarray}
  for the jog located at $z=z_m$.
  The first term in these formulas come from the pipe diffusion vacancy flux which is along the dislocation, and the second terms come from the bulk diffusion vacancy flux which is in the plane normal to the dislocation.
As in the above diffusion equations, when the solution of the bulk diffusion $c^v$ is symmetric with respect to $r$, this jog velocity is reduced to $v_{\rm jog}^{(m)}=D_c\big[c^c_z(z_m^+)-c^c_z(z_m^-)\big]+\frac{2\pi r_dD_v}{bl_\phi}\big(c^v-k_v c_0^ce^{-\frac{f_{\rm cl}\Omega}{bkT}}\big)|_{z=z_m}$.

\section{The dislocation dynamics model}\label{sec:IV}

 \begin{figure}[htbp]
 \centering
 \includegraphics[width=15cm]{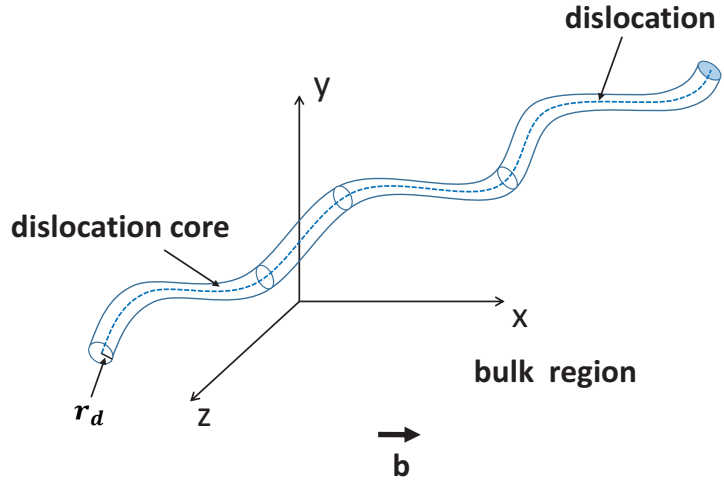}
  \caption{The dislocation dynamics model for dislocation climb.}
  \label{fig:dislocationdynamics}
\end{figure}

In this section,  we derive a  dislocation dynamics model for vacancy-assisted dislocation climb by further upscaling in space and time from the jog dynamics model obtained in the previous section. We assume that:

\begin{enumerate}[(i)]
      \item
 The jogs on dislocations are sparsely distributed on the atomic length scale, and the average distance between two adjacent jogs is much smaller than the length scale of the dislocation dynamics model. That is, if we denote the average distance between two adjacent jogs on the dislocation by $l$ and the domain size of the dislocation dynamics model by $L$, this assumption can be written as $b<<l<<L$.

 \item
 The vacancy pipe diffusion is much faster than the vacancy bulk diffusion. That is, $D_c>>D_v$.

  \item
  It takes longer time to emit a vacancy from dislocation core to the bulk than to diffuse a vacancy from one jog to an adjacent jog along the dislocation core. That is, $\sqrt{D_c\tau_e}>>l$, where $\tau_e=\frac{b^3}{2\pi r_d D_v\phi_vk_v}=\frac{b^2l_\phi c^c_0}{2\pi r_d D_vc_0}$ is the time to emit one vacancy from the dislocation core to the bulk.

  \item
  In the dislocation dynamics model, a dislocation is a smooth curve without explicit jog structure,  see Fig.~\ref{fig:dislocationdynamics}, and the effects of jogs and pipe diffusion are included in the formulation of its dynamics by the following averaging process. At a jog $z_m$ on the dislocation, the climb velocity  $v_{\rm cl}$, the boundary condition   for the bulk vacancy diffusion and the equilibrium vacancy  concentration  $c^c$  are calculated by local averages over the portion of $[z_m-\frac{l_{m-1}}{2},z_{m}+\frac{l_m}{2}]$ on the dislocation, where $l_m=z_{m+1}-z_m$. The values of these quantities  are then extended to the entire dislocation by connecting smoothly their values on the jogs.

  \end{enumerate}

Consider the vacancy pipe diffusion along the dislocation segment between the two adjacent jogs at $z_m$ and $z_{m+1}$. Under the fast pipe diffusion in Assumption (ii) above, we can further  assume pipe diffusion equilibrium in Eq.~\eqref{eq-core-diffusion-equation}, which gives
\begin{eqnarray}\label{eq-pipe-diffusion-equation}
    \left\{
        \begin{array}{l}
            D_cc^c_{zz}-\frac{1}{\tau_e}c^c+F(z)=0, \ z\in (z_m,z_{m+1}),\\
            c^c(z_m)=c_J (z_m), \\
                        c^c(z_{m+1})=c_J(z_{m+1}),
        \end{array}
    \right.
\end{eqnarray}
where  $F(z)=\frac{D_v}{b^2l_\phi} \int_{r=r_d} c^v(x,y,z) \ dl$ is the vacancy flux from the bulk into the dislocation core, and recall that
$\tau_e$ defined in Assumption (iii) is the time to emit one vacancy from the dislocation core to the bulk and  $c_J(z_i)=c^c_0e^{-\frac{f_{\rm cl}\Omega}{bkT}}|_{z=z_i}$ is the equilibrium vacancy concentration at the jog $z_i$.

Given the vacancy concentration $c^v$ in the bulk, or the function $F(z)$, the above ODE boundary value problem has the following analytical solution
\begin{eqnarray}\label{eq-core-diffusion-solution}
    c^c(z)&=&\left(C_m^--\frac{1}{2}\sqrt{\frac{\tau_e}{D_c}}
    \int_{z_m}^{z}F(\omega)e^{-\frac{1}{\sqrt{D_c\tau_e}}\omega}\D \omega\right)e^{\frac{1}{\sqrt{D_c\tau_e}}z}\nonumber\\
    &&+\left(C_m^++\frac{1}{2}\sqrt{\frac{\tau_e}{D_c}}
    \int_{z_m}^{z}F(\omega)e^{\frac{1}{\sqrt{D_c\tau_e}}\omega}\D \omega\right)e^{-\frac{1}{\sqrt{D_c\tau_e}}z},
\end{eqnarray}
where
\begin{eqnarray}
    C_m^-&=&
    \frac{-c_J(z_{m+1}) e^{-\frac{z_m}{\sqrt{D_c\tau_e}}}+
   c_J(z_m)e^{-\frac{z_{m+1}}{\sqrt{D_c\tau_e}}}-\mathcal{F}_m^- e^{\frac{z_{m+1}-z_m}{\sqrt{D_c\tau_e}}}
   +\mathcal{F}_m^+ e^{-\frac{z_m+z_{m+1}}{\sqrt{D_c\tau_e}}} }{ e^{\frac{z_m-z_{m+1}}{\sqrt{D_c\tau_e}}}-  e^{\frac{z_{m+1}-z_m}{\sqrt{D_c\tau_e}}}}
    \label{eq..coefficient1}
   \\
    C_m^+&=&
    \frac{c_J(z_{m+1}) e^{\frac{z_m}{\sqrt{D_c\tau_e}}}-c_J(z_m)
   e^{\frac{z_{m+1}}{\sqrt{D_c\tau_e}}}-\mathcal{F}_m^+ e^{\frac{z_m-z_{m+1}}{\sqrt{D_c\tau_e}}}
   +\mathcal{F}_m^- e^{\frac{z_m+z_{m+1}}{\sqrt{D_c\tau_e}}}}{ e^{\frac{z_m-z_{m+1}}{\sqrt{D_c\tau_e}}}- e^{\frac{z_{m+1}-z_m}{\sqrt{D_c\tau_e}}}}
    \label{eq..coefficient2}
\end{eqnarray}
with
\begin{eqnarray}\label{eq..integral.F1}
    \mathcal{F}_m^-&=&\frac{1}{2}\sqrt{\frac{\tau_e}{D_c}}
    \int_{z_m}^{z_{m+1}}F(\omega)e^{-\frac{1}{\sqrt{D_c\tau_e}}\omega}\D \omega,\\
        \mathcal{F}_m^+&=&\frac{1}{2}\sqrt{\frac{\tau_e}{D_c}}
        \int_{z_m}^{z_{m+1}}F(\omega)e^{\frac{1}{\sqrt{D_c\tau_e}}\omega}\D \omega.\label{eq..integral.F2}
\end{eqnarray}
This is the equilibrium vacancy concentration on the dislocation segment $[z_m,z_{m+1}]$ for any integer $m$.

Now we consider the velocity of the jog $z_m$ given by Eq.~\eqref{velocity-jog2}. The first term in the expression is the contribution from the vacancy flux into this jog from $z_m^+$ side depending on the equilibrium vacancy concentration on the dislocation segment $[z_m,z_{m+1}]$, and the vacancy flux into it from $z_m^-$ side depending on the equilibrium vacancy concentration on  $[z_{m-1},z_{m}]$. Using the above solution formula, this contribution from pipe diffusion is
\begin{eqnarray}
&&\mathbf j^c\cdot \mathbf n^c|_{z=z_m^+}+\mathbf j^c\cdot \mathbf n^c|_{z=z_m^-}\nonumber\\
 & =&D_cc^c_z(z_m^+)-D_cc^c_z(z_m^-)\vspace{1ex}\nonumber\\
 & =&\sqrt{\frac{D_c}{\tau_e}}\left[C_m^-e^{\frac{1}{\sqrt{D_c\tau_e}}z_m}
  -C_m^+e^{-\frac{1}{\sqrt{D_c\tau_e}}z_m}\right]\nonumber\\
&&  -\sqrt{\frac{D_c}{\tau_e}}\left[(C_{m-1}^--\mathcal{F}_{m-1}^-)e^{\frac{1}{\sqrt{D_c\tau_e}}z_m}
  -(C_{m-1}^++\mathcal{F}_{m-1}^+)e^{-\frac{1}{\sqrt{D_c\tau_e}}z_m}\right].\label{eqn:jc}
\end{eqnarray}

Using Eq.~\eqref{eq..integral.F1} and $l_m=z_{m+1}-z_m<<\sqrt{D_c\tau_e}$ by Assumption (iii), we have
\begin{eqnarray}
\mathcal{F}_m^- e^{\frac{z_{m+1}}{\sqrt{D_c\tau_e}}}&=&\frac{1}{2}\sqrt{\frac{\tau_e}{D_c}}\left(
    \int_{z_m}^{z_{m+1}}F(\omega)e^{-\frac{1}{\sqrt{D_c\tau_e}}\omega}\D \omega\right)e^{\frac{z_{m+1}}{\sqrt{D_c\tau_e}}}\nonumber\\
    &\approx& \frac{1}{2}\sqrt{\frac{\tau_e}{D_c}}e^{\frac{z_{m+1}}{\sqrt{D_c\tau_e}}}
    \int_{z_m}^{z_{m+1}}\big[F(z_m)+F'(z_m)(w-z_m)\big]e^{-\frac{1}{\sqrt{D_c\tau_e}}\omega}\D \omega \nonumber\\
    &=&\frac{1}{2}\sqrt{\frac{\tau_e}{D_c}}\left[ F(z_m)\sqrt{D_c\tau_e}  \left(e^{\frac{z_{m+1}-z_m}{\sqrt{D_c\tau_e}}}-1\right)-F'(z_m)l_m \sqrt{D_c\tau_e} \right]\nonumber\\
    &&+F'(z_m) D_c\tau_e  \left(e^{\frac{z_{m+1}-z_m}{\sqrt{D_c\tau_e}}}-1\right)\nonumber\\
    &\approx&\frac{1}{2}\sqrt{\frac{\tau_e}{D_c}}\left[ F(z_m)l_m
    +\frac{1}{2\sqrt{D_c\tau_e}}\left(F(z_m)+F'(z_m)\sqrt{D_c\tau_e}\right)l_m^2  \right].
\end{eqnarray}
Similarly, using Eq.~\eqref{eq..integral.F2}, we have
\begin{eqnarray}
\mathcal{F}_m^+ e^{-\frac{z_{m+1}}{\sqrt{D_c\tau_e}}}    \approx\frac{1}{2}\sqrt{\frac{\tau_e}{D_c}}\left[ F(z_m)l_m
+\frac{1}{2\sqrt{D_c\tau_e}}\left(-F(z_m)+F'(z_m)\sqrt{D_c\tau_e}\right)l_m^2  \right].
\end{eqnarray}
Using Eqs.~\eqref{eq..coefficient1} and \eqref{eq..coefficient2}, together with the above approximations and the Taylor expansion of $c_J(z)$ at $z_m$,  we have
\begin{eqnarray}
&& C_m^-e^{\frac{1}{\sqrt{D_c\tau_e}}z_m}-C_m^+e^{-\frac{1}{\sqrt{D_c\tau_e}}z_m}\nonumber\\
 &=&\sqrt{D_c\tau_e}c_J'(z_m)+\frac{1}{2}\sqrt{D_c\tau_e}
  \left(\frac{1}{D_c}F(z_m)-\frac{c_J(z_m)}{D_c\tau_e}+c_J''(z_m)\right)l_m
  +O(l_m^2).
\end{eqnarray}

Similarly, we have
\begin{eqnarray}
&& (C_{m-1}^--\mathcal{F}_{m-1}^-)e^{\frac{1}{\sqrt{D_c\tau_e}}z_m}
  -(C_{m-1}^++\mathcal{F}_{m-1}^+)e^{-\frac{1}{\sqrt{D_c\tau_e}}z_m}\nonumber\\
 &=&\sqrt{D_c\tau_e}c_J'(z_m)+\frac{1}{2}\sqrt{D_c\tau_e}
  \left(-\frac{1}{D_c}F(z_m)+\frac{c_J(z_m)}{D_c\tau_e}-c_J''(z_m)\right)l_{m-1}
  +O(l_{m-1}^2).
\end{eqnarray}
Thus from Eq.~\eqref{eqn:jc}, we have
\begin{eqnarray}\label{eqn:jc1}
\mathbf j^c\cdot \mathbf n^c|_{z=z_m^+}+\mathbf j^c\cdot \mathbf n^c|_{z=z_m^-}
\approx  \left(F(z_m)-\frac{c_J(z_m)}{\tau_e}+D_cc_J''(z_m)\right)\frac{l_m+l_{m-1}}{2}.
  \end{eqnarray}

Moreover, the contribution to the velocity of $z_m$ in Eq.~\eqref{velocity-jog2} due to the bulk diffusion vacancy flux into this jog is
  \begin{eqnarray}\label{eqn:jc2}
  \left.\frac{1}{b}\int_{r=r_d}\mathbf j\cdot \mathbf n \ dl\right|_{z=z_m}
  &=&\left.\frac{2\pi r_d D_v}{bl_\phi}\left(\frac{1}{2\pi r_d}\int_{r=r_d} c^v \ dl-k_v c^c_0e^{-\frac{f_{\rm cl}\Omega}{bkT}}\right)\right|_{z=z_m}\nonumber\\
  &=&\left(F(z_m)-\frac{c_J(z_m)}{\tau_e}\right)b.
  \end{eqnarray}
Since $l_m, l_{m-1}>>b$, this contribution to the jog velocity from the  bulk diffusion vacancy flux is small compared with the contribution from the pipe diffusion in Eq.~\eqref{eqn:jc1}. Therefore, we have the following jog velocity formula
\begin{eqnarray}\label{velocity-jog-final}
  v_{\rm jog}^{(m)}
  &=&\left(F(z_m)-\frac{c_J(z_m)}{\tau_e}+D_cc_J''(z_m)\right)\frac{l_m+l_{m-1}}{2}\nonumber\\
    &=&\left( \frac{D_v}{b^2l_\phi} \int_{r=r_d} c^v(x,y,z_m) \ dl-\frac{c_J(z_m)}{\tau_e}+D_cc_J''(z_m)\right)\frac{l_m+l_{m-1}}{2}.
  \end{eqnarray}

The local dislocation climb velocity can be calculated from the jog velocity by $v_{\rm cl}(z_m)=\frac{b}{(l_m+l_{m-1})/2}v_{\rm jog}^{(m)}$, which is
\begin{eqnarray}\label{eqn:climbatjog}
v_{\rm cl}(z_m)=\left( \frac{D_v\phi_v}{b^3} \int_{r=r_d} c^v(x,y,z_m) \ dl
-\frac{c_J(z_m)}{\tau_e}+D_cc_J''(z_m)\right)b.
\end{eqnarray}

Now we perform the averaging process as described in  Assumption (iv). First,
the average of $c^c(z)$ over $[z_m-\frac{l_{m-1}}{2},z_{m}+\frac{l_m}{2}]$, which from the pipe diffusion equation is
\begin{eqnarray}
c^c_d(z_m)&=&\frac{2}{l_m+l_{m-1}}\int^{z_{m}+\frac{l_m}{2}}_{z_m-\frac{l_{m-1}}{2}} c^c(z)\D z\nonumber\\
&=&\frac{2\tau_e}{l_m+l_{m-1}} \int^{z_{m}+\frac{l_m}{2}}_{z_m-\frac{l_{m-1}}{2}}F(z)\D z
-\frac{2\tau_e}{l_m+l_{m-1}} \big[D_cc^c_z(z_m^+)-D_cc^c_z(z_m^-)\big]\nonumber\\
&&+\frac{2\tau_e}{l_m+l_{m-1}} \left[D_cc^c_z(z_m+\frac{l_m}{2})-D_cc^c_z(z_m-\frac{l_{m-1}}{2})\right]
\vspace{1ex}\nonumber\\
&\approx&\tau_e F(z_m)-\tau_e \left(F(z_m)-\frac{c_J(z_m)}{\tau_e}+D_cc_J''(z_m)\right)+\tau_eD_cc_J''(z_m)\vspace{1ex}\nonumber\\
&=&c_J(z_m).
\end{eqnarray}
Thus after extending the values to the entire dislocation, we have  the average equilibrium vacancy concentration at a point on the dislocation
\begin{eqnarray}\label{eqn:c-c-average-0}
c^c_d(z)=c_0^ce^{-\frac{f_{\rm cl}(z)\Omega}{bkT}}.
\end{eqnarray}

Next, averaging over $[z_m-\frac{l_{m-1}}{2},z_{m}+\frac{l_m}{2}]$ along the dislocation the Robin boundary condition in the vacancy bulk diffusion problem given in Eq.~\eqref{eq-diffusion-equation} in the previous section, we have
\begin{eqnarray}
 -\frac{\partial c}{\partial n}
 =\frac{1}{l_\phi}\big(c-k_v c^c_d\big)=\frac{1}{l_\phi}\big(c-c_d\big),
\end{eqnarray}
where $c$ is the solution of the vacancy bulk diffusion equation corresponding to this averaged boundary condition, and $c_d(z)$ is defined by $c_d(z)=k_vc^c_d(z)$ and can be written as
\begin{eqnarray}\label{eqn:c-c-average}
c_d(z)=c_0e^{-\frac{f_{\rm cl}(z)\Omega}{bkT}}.
\end{eqnarray}
 Here we have used the relation $k_vc_0^c=c_0$ due to detailed balance, where $c_0$ is recalled to be the reference equilibrium vacancy concentration in the bulk. See the discussion at the end of Sec.~2.1. The vacancy concentration $c_d(z)$ is in fact the equilibrium vacancy concentration when approaching the dislocation core from the bulk.

Therefore, after averaging, the vacancy bulk diffusion equation in Eq.~\eqref{eq-diffusion-equation} becomes
\begin{eqnarray}\label{eq-diffusion-equation-av}
    \left\{
        \begin{array}{l}
     c_t= D_v \big( c_{xx}+ c_{yy}+ c_{zz}\big), \ {\rm in \ the \ bulk},\vspace{1ex}\\
          {\displaystyle -\frac{\partial c}{\partial n} =\left.\frac{1}{l_\phi}\big(c-c_d\big)\right|_{r=r_d}},  \vspace{1ex}\\
          c=\left. c_{\infty} \right|_{r=r_\infty},
        \end{array}
    \right.
\end{eqnarray}
and the climb velocity in Eq.~\eqref{eqn:climbatjog} becomes
\begin{eqnarray}\label{eqn:climb-av}
v_{\rm cl}&=&
\frac{2\pi r_dD_v }{bl_\phi}\left(\frac{1}{2\pi r_d}\int_{r=r_d} c \ dl-c_d\right)
+D_cb\frac{\D^2 c^c_d}{\D s^2}\vspace{1ex}\\
&=&\frac{1}{b}\int_{r=r_d}\mathbf j\cdot \mathbf n \ dl+D_cb\frac{\D^2 c^c_d}{\D s^2} \label{eqn:climb-av2}
\end{eqnarray}
at any point on the dislocation, where $c_d$ and $c^c_d$ are the equilibrium vacancy concentrations on the dislocation core surface from outside and inside given by Eqs.~\eqref{eqn:c-c-average} and \eqref{eqn:c-c-average-0}, respectively. Here we have generalized the formulation to curved dislocation with $s$ being the arc-length parameter along the dislocation.

In summary, the vacancy-assisted dislocation climb model for dislocation dynamics simulation includes the climb velocity given in Eq.~\eqref{eqn:climb-av2} and the boundary value problem of vacancy bulk diffusion in Eq.~\eqref{eq-diffusion-equation-av}, where the equilibrium vacancy concentrations on the dislocation core surface from outside and inside,  $c_d$ and $c^c_d$,
 are given by Eqs.~\eqref{eqn:c-c-average} and \eqref{eqn:c-c-average-0}, respectively. In this formulation,  the climb velocity $v_{\rm cl}$ in Eq.~\eqref{eqn:climb-av2}  has contributions  from both the net vacancy flux from the bulk into the dislocation (the first term, depending on $D_v$), and the contribution that accounts for the effect of pipe diffusion and stress variation along the dislocation (the second term, depending on $D_c$ and $c^c_d$). Moreover, in this formulation, a Robin boundary condition is used on the dislocation core surface for the vacancy bulk diffusion in Eq.~\eqref{eq-diffusion-equation-av}, meaning the role of an imperfect vacancy sink/source served by a dislocation due to the relatively slow exchange of vacancies between the dislocation core and the bulk compared with the bulk diffusion.     Interestingly, the derived Robin boundary condition and climb velocity as leading order approximations do not depend on the lengths of dislocation segments between jogs.
{ The parameter $l_\phi$ in the Robin boundary condition is a characteristic length that is associated with the difference between the barrier for the vacancies hopping into the dislocation core from the bulk  $E^{v-c}$ and that for the vacancy bulk diffusion $E^v$, and is given by $l_\phi=b/\phi_v= b e^{\frac{ E^{v-c}-E^v}{kT}}$, if we assume the temperature-independent prefactors in the two hopping rates are the same. Also note that  the equilibrium vacancy concentration  on the dislocation core surface from outside $c_d$  appears in the boundary condition of the vacancy bulk diffusion equation, and the equilibrium vacancy concentration inside the dislocation core $c^c_d$ appears in the pipe diffusion contribution in the climb velocity expression. }

Compared with the classical vacancy-assisted dislocation climb model \citep{Hirth-Lothe}, this new formulation includes the effects of vacancy pipe diffusion and absorption/emission on jogged  dislocations and the vacancy exchange between dislocation core and the bulk, by the new term of second derivative of vacancy concentration along the dislocation in the climb velocity and the Robin boundary condition (meaning imperfect vacancy sink/source) instead of the Dirichlet boundary condition (i.e. $c=c_d$ on the dislocation core surface $r=r_d$, meaning perfect vacancy sink/source) in the classical model.
{ In the limit cases of $l_\phi<<r_d$, meaning the barrier for the vacancies hopping into the dislocation core from the bulk is appreciably smaller than the barrier for the  vacancy bulk diffusion,  our
Robin boundary condition reduces to the Dirichlet boundary condition in the classical  dislocation climb model; otherwise, i.e. when the barrier for the vacancies hopping into the dislocation core from the bulk is comparable or greater than the barrier for the  vacancy bulk diffusion, the Robin boundary condition should be used.}
The new term of second derivative of vacancy concentration along the dislocation core in the climb velocity has never been  reported in the available models in the literature to our knowledge. This new term is able to explain the translation of prismatic loops at low temperatures observed in experiments \citep{Hirth-Lothe,Kroupa1961}, as will discussed in more details in the next section.

 A Robin boundary condition in its form in Eq.~\eqref{eq-diffusion-equation-av} was adopted in a phenomenological way in the two dimensional dislocation climb model in \cite{Ayas2014113}. Our Robin boundary condition is derived rigorously by upscaling from the  microscopic models  and incorporates all the four microscopic mechanisms given in the introduction section, under the condition of fast vacancy pipe diffusion. This result applies to both the diffusion-limited and  sink-limited cases considered by \cite{Ayas2014113} under the fast vacancy pipe diffusion condition.
 Our upscaling method in principle can also be employed to rigorously derive the boundary condition in their sink-limited case  without fast pipe diffusion, in which the jogs are far apart beyond the mean  pipe diffusion length of a vacancy absorbed from the bulk and each jog acts as an isolated sink of vacancies. This will be explored in the future work.

{The parameters in the derived vacancy-assisted dislocation climb model for dislocation dynamics simulations in Eqs.~\eqref{eqn:climb-av2}, \eqref{eq-diffusion-equation-av}, and Eq.~\eqref{eqn:c-c-average}, \eqref{eqn:c-c-average-0} can be calculated by atomistic models~\citep[e.g.][]{Hoagland1998,Fang2000,Mishin2009,Yip2009,Yip2010,YZWang2012,Dudarev2016}. Despite many available atomistic results  for the other parameters in the model associated with vacancy bulk and pipe diffusion, few  atomistic results are available for the new parameter $l_\phi$ associated with the energy barriers $E^{v-c}$  for the  vacancies hopping into the dislocation core from the bulk.
 \citet{Yip2009} and \citet{Yip2010} have calculated such energy barriers for the $<111>(110)$ $71^\circ$ edge-type dislocation in bcc iron, when a vacancy hops from the bulk to the center of dislocation core along different paths. The values they obtained, for example,
 are $E^{v-c}-E^v=-0.35eV$ and $0.04eV$ along their paths A and  D, respectively (read  from  Fig.~1 in \citet{Yip2010} and averaged within the dislocation core whose radius is $r_d=4b$).  These give $l_\phi=0.017b$  along path A and $l_\phi=1.6b$  along path D at $T=1000K$.  In the former case, $l_\phi<<b$ and the classical Dirichlet boundary condition can be used for the vacancy bulk diffusion problem. Whereas in the latter case, $l_\phi$ is comparable with the core radius $r_d=4b$ and the Robin boundary condition should be used. These atomistic results indicate a wide range of possible values of $l_\phi$ and the Robin boundary condition should in general be employed  for the vacancy bulk diffusion problem. For these $<111>(110)$ $71^\circ$ edge-type dislocations in bcc iron, the actual value of $l_\phi$ used in the vacancy bulk diffusion problem should be some average along these paths. Such averaging and systematic atomistic calculations of $l_\phi$ for other materials will be explored in the future work.}

In many cases, the dislocation climb process is much slower than the vacancy bulk diffusion. In these cases, vacancy bulk diffusion  equilibrium can be assumed \citep{Hirth-Lothe}, and the equilibrium vacancy concentration in the bulk can be obtained by solving the following diffusion equilibrium problem
\begin{eqnarray}\label{eq-diffusion-equilibrium}
    \left\{
        \begin{array}{l}
     D_v \big( c_{xx}+ c_{yy}+ c_{zz}\big)=0, \ {\rm in \ the \ bulk},\vspace{1ex}\\
          {\displaystyle - \frac{\partial c}{\partial n} =\left.\frac{1}{l_\phi}\big(c-c_d\big)\right|_{r=r_d}},  \vspace{1ex}\\
          c=\left. c_{\infty} \right|_{r=r_\infty}.
        \end{array}
    \right.
\end{eqnarray}
The climb velocity $v_{\rm cl}$ is still given by Eq.~\eqref{eqn:climb-av2} or \eqref{eqn:climb-av}.

\section{Dislocation climb velocities of some special cases}\label{sec:V}

Let us apply our dislocation dynamics formulation for vacancy-assisted dislocation climb derived in the previous section to some specific cases which are of practical interests. This also helps understand the physical interpretation of our formulation.

\subsection{Climb velocity of a straight edge dislocation}
We consider a straight edge  dislocation along the $z$ axis with Burgers vector $\mathbf b$ in the $+x$ direction that climbs in the $y$ direction under a constant applied stress. In this case, all the quantities are uniform along the dislocation. Further under the equilibrium assumption for the  vacancy bulk diffusion,
 the equilibrium diffusion equation in Eq.~\eqref{eq-diffusion-equilibrium} is reduced to
\begin{eqnarray}
c_{xx}+c_{yy}=0.
\end{eqnarray}
The solution of this equation with the boundary conditions in Eq.~\eqref{eq-diffusion-equilibrium}
can be written in the form
\begin{eqnarray}
c(r)=Q_1\ln r +Q_2,
\end{eqnarray}
where $r=\sqrt{x^2+y^2}$ is the distance to the dislocation, and $Q_1$ and $Q_2$ are two constants.
Using the boundary conditions at $r=r_d$ and $r=r_\infty$ in Eq.~\eqref{eq-diffusion-equilibrium}, we have
\begin{eqnarray}
c(r)=\frac{c_\infty- c_d}{\ln \frac{r_\infty}{r_d}+\frac{l_\phi}{r_d}}\ln \frac{r}{r_\infty}+c_\infty.
\end{eqnarray}
By the climb velocity formula in Eq.~\eqref{eqn:climb-av}, we have
\begin{eqnarray}\label{eq-velocity-edge}
v_{\rm cl}=\frac{2\pi D_v}{b\left(\ln \frac{r_\infty}{r_d}+\frac{l_\phi}{r_d}\right)}\big(c_{\infty}-c_d\big).
\end{eqnarray}

Comparing with the classical climb formula for a straight edge dislocation \cite{Hirth-Lothe}:
  \begin{eqnarray}\label{eq-velocity-edge0}
v_{\rm cl}=\frac{2\pi D_v }{b\ln\frac{ r_{\infty}}{r_d}}\big(c_{\infty}-c_d\big),
\end{eqnarray}
which is obtained by using the Dirichlet boundary condition $c=c_d$ on the dislocation core surface $r=r_d$,
it can be seen that our climb formula has an extra term $l_\phi/ r_d$ in the denominator, which is due to the role of imperfect sink/source represented by the Robin boundary condition in our model that results from the slow vacancy exchange between the dislocation core and the bulk compared with the bulk diffusion, leading to a smaller climb velocity compared with the classical formula using Dirichlet boundary condition. In the limit cases of $l_\phi<<r_d$ where the barrier for the vacancies hopping into the dislocation core from the bulk is smaller than the barrier for the  vacancy bulk diffusion,  our climb velocity formula in Eq.~\eqref{eq-velocity-edge}
 also reduces to the classical  one in Eq.~\eqref{eq-velocity-edge0}, as the boundary condition on the dislocation core surface for vacancy bulk diffusion discussed in the previous section.

Recently \citet{Pierre-Antoine} derived a climb velocity formulation for a single straight dislocation by solving the coupled system of vacancy pipe diffusion and bulk diffusion equations (their Eq.~(1)). Our climb velocity formula in Eq.~\eqref{eq-velocity-edge} is consistent with their formula in the limit case of faster pipe diffusion than the exchange of vacancies between dislocation core and the bulk, i.e. $\sqrt{D_c\tau_e}>>l$  as in our Assumption (iii) (or $l_c>>d_j$ in their notations).

\subsection{Climb velocity of a circular prismatic loop}
\begin{figure}[htbp]
\centering
\includegraphics [width=10cm]{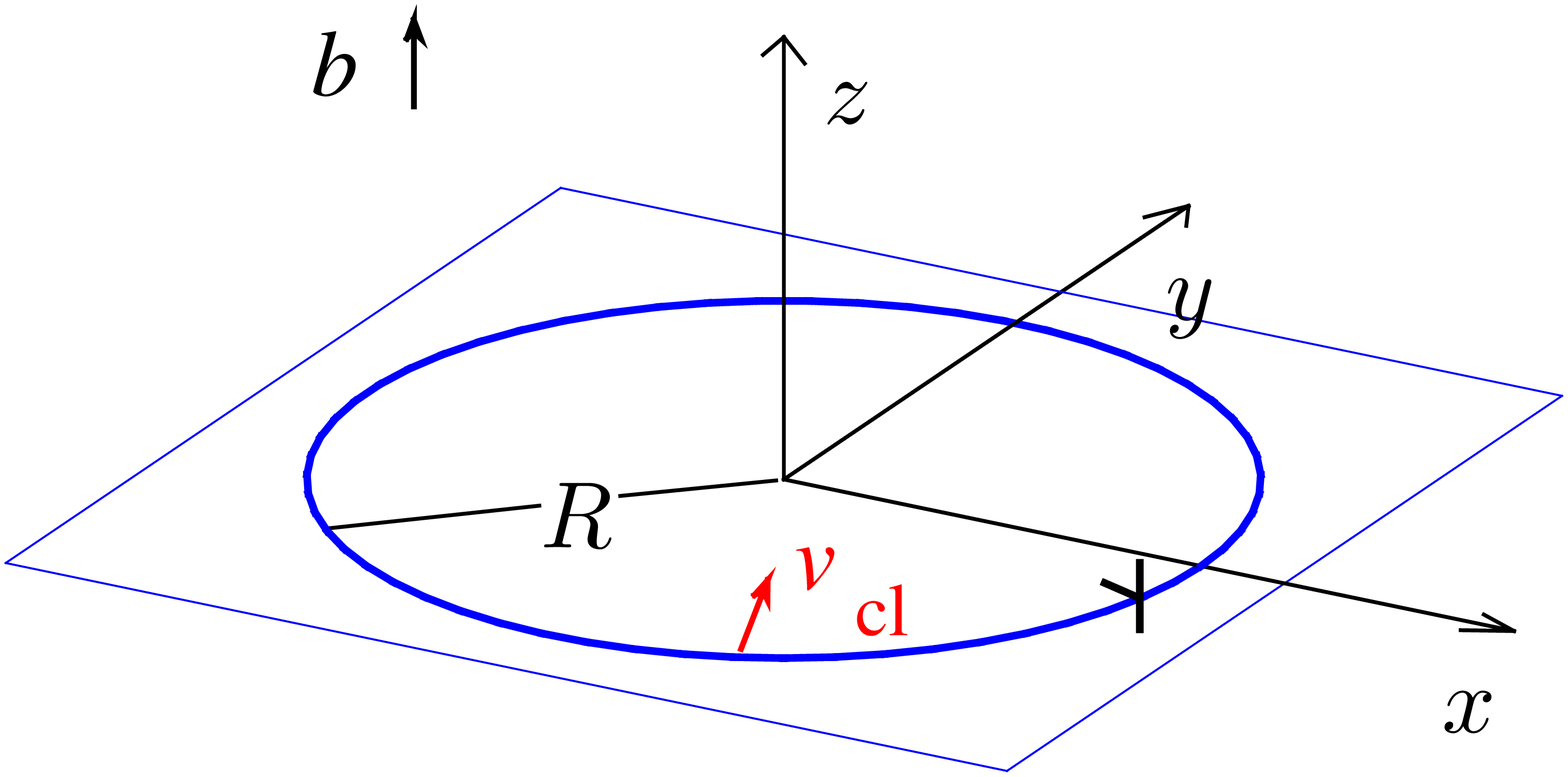}
\caption{Climb of a circular prismatic loop.}
\label{fig:singleloop}
\end{figure}

Next, we consider the climb of a circular prismatic dislocation loop, see Fig.~\ref{fig:singleloop}.
The circular loop has radius $R$ located in the $xy$ plane centered at the origin. The Burgers vector  $\mathbf b$ of this loop is in the $z$ direction.
The loop shrinks by climb driven by a  Peach-Koehler force arising from its  self-stress. Due to symmetry, the climb force and velocity are  constants over the loop. The Peach-Koehler force on the dislocation can be calculated \citep{Hirth-Lothe,Gu2015319} as $f_{\rm cl}(x)=\frac{ \mu b^2}{4\pi(1-\nu)R} \left(\ln\frac{8R}{r_d}-1\right)$, which is
 in the inward radial direction.

Following \citet{Gu2015319}, we use
the Green's function formulation to express the solution of the vacancy bulk diffusion equation in Eq.~\eqref{eq-diffusion-equilibrium} as
\begin{equation}
c(x,y,z)=-\frac{bv_{\rm cl}R}{4\pi D_v}\int_0^{2\pi}{\frac{\ud\theta}
{\sqrt{(x-R\cos\theta)^2+(y-R\sin\theta)^2+z^2}}}+c_\infty.
\end{equation}
This solution satisfies the last boundary condition  $c=c_{\infty}$ at the far field. We consider the boundary condition at the point $(R,0,r_d)$ on the dislocation core surface, which is
\begin{align}\label{green-robin}
&-\frac{bv_{\rm cl}R}{4\pi D_v}\int_0^{2\pi}{\frac{r_d}
{\big[R+r_d-R\cos\theta)^2+(R\sin\theta)^2\big]^\frac{3}{2}}}\ud\theta\nonumber\\
=&\frac{1}{l_\phi}\left[-\frac{bv_{\rm cl}R}{4\pi D_v}\int_0^{2\pi}{\frac{\ud\theta}
{\sqrt{(R+r_d-R\cos\theta)^2+(R\sin\theta)^2}}}+c_\infty
-c_0e^{-\frac{f_{\rm cl}\Omega}{bk_BT}}\right].
\end{align}
Since $R>>r_d$, using the complete elliptic integrals of the first kind $K(k)$ and
 the asymptotic behavior $K(k)\sim \ln(4/\sqrt{1-k^2})$ as $k\rightarrow 1^-$, it can be calculated that
 \begin{equation}
v_{\rm cl}=\frac{2\pi D_v}{b\left(\ln\frac{8R}{r_d}+\frac{l_\phi}{r_d}\right)}
\left(c_\infty-c_0e^{-\frac{f_{\rm cl}\Omega}{bk_BT}}\right).
\end{equation}
Compared with the available results \citep{Hirth-Lothe,Gu2015319}, as the climb velocity of a straight edge dislocation obtained in the previous subsection, there is also an additional contribution $l_\phi/r_d$ in the denominator due to the Robin boundary condition. The climb velocity is also smaller than the classical formula using the Dirichlet boundary condition.

If we set $c_\infty=c_0$ and for $f_{\rm cl}<<b k_BT/\Omega$, and
inserting the Peach-Koehler force  into this result yields the climb velocity of a circular prismatic loop under its own self-stress:
\begin{equation}
v_{\rm cl}\approx\frac{ \mu D_vc_0\Omega}{2(1-\nu)k_BTR(1+l_\phi/r_d)}.
\end{equation}

\subsection{Prismatic loop translation by pipe diffusion}

\begin{figure}[htbp]
 \centering
 \subfigure[]
 {\includegraphics[width=10cm]{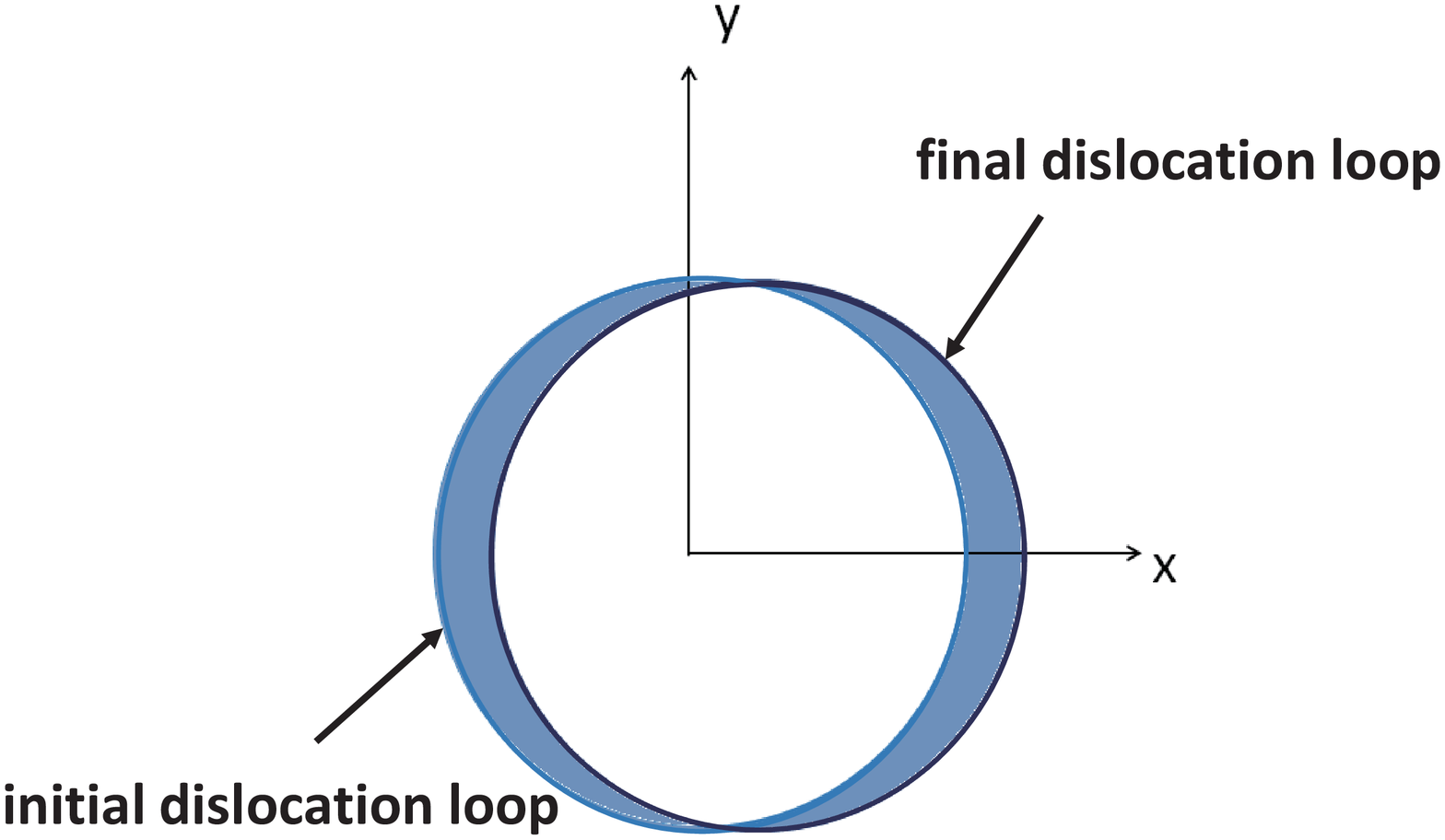}}
 \subfigure[]
 {\includegraphics[width=6cm]{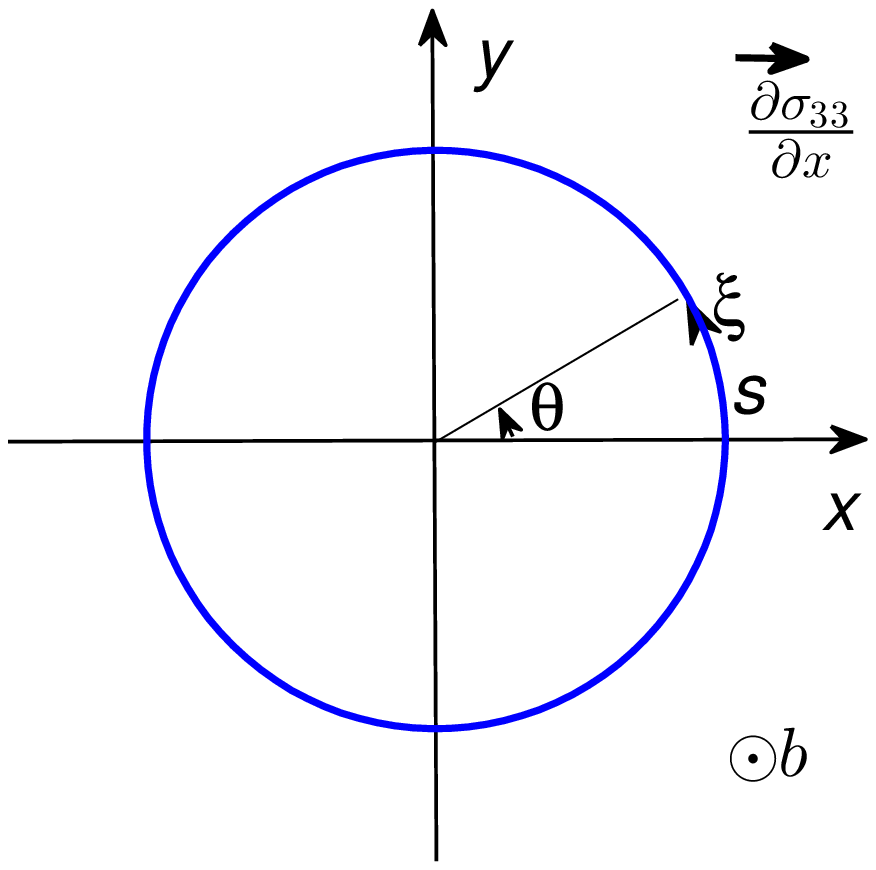}}
  \caption{(a) Translation of a prismatic loop at low temperatures. (b) Parametrization of the circular prismatic loop.}
  \label{fig:translation}
\end{figure}

Finally, we consider a prismatic loop $\gamma$ at a low temperature. In this case, vacancy bulk diffusion is negligible, i.e. $D_v\approx0$, and only the vacancies within the dislocation core can move by pipe diffusion. It  has been observed in experiments \citep{Hirth-Lothe,Kroupa1961} that the loop can translate under a stress gradient, see an illustration in Fig.~\ref{fig:translation}(a).

Under this condition, our climb velocity formula in Eq.~\eqref{eqn:climb-av2} is reduced to
\begin{eqnarray}\label{eqn:climb-dv0}
v_{\rm cl}=D_cb\frac{\D^2 c^c_d}{\D s^2},
\end{eqnarray}
where $s$ is recalled to be the arclength of the loop. By Eq.~\eqref{eqn:c-c-average}.
the vacancy concentration in the dislocation core $c^c_d$ depends on the climb force or the stress field as
\begin{eqnarray}\label{eqn:climb-dv0-1}
c^c_d(s)=c^c_0e^{-\frac{f_{\rm cl}(s)\Omega}{bkT}}.
\end{eqnarray}
This means that the loop can move under the variation of stress along the dislocation. { Note that in this case, vacancies are emitted by the jogs on the loop, and these emitted vacancies move along the dislocation core by pipe diffusion. For this prismatic loop $\gamma$ under appropriate stress field, the jogs on one half of the loop emit vacancies, while those on the other half-loop absorb vacancies. As a result, the entire loop moves. }

For this prismatic loop $\gamma$ moving with the above velocity, the rate of change of the area $S$ enclosed by the loop is
\begin{eqnarray}\label{eqn:climb-dv1}
\frac{\D S}{\D t}=\int_\gamma v_{\rm cl}\D s =D_cb\int_\gamma \frac{\D^2 c^c_d}{\D s^2}\D s=0.
\end{eqnarray}
This shows that using our formulation, the area $S$ that is enclosed by the loop is unchanged by the translation due to vacancy pipe diffusion, which agrees with the experimental observations. { The shape of the prismatic loop during the translation depends on the nature of the stress field.}

Now we examine the climb velocity formula in Eq.~\eqref{eqn:climb-dv0} for a circular prismatic loop. Suppose that the
circular prismatic loop is centered at the origin and has radius $R$, and is parametrized by the polar coordinates as $x=R\cos\theta$, $y=R\sin\theta$, see Fig.~\ref{fig:translation}(b). Suppose the stress field vary with $x$ as $\sigma_{33}=\sigma_{33}(x)$.
 Thus both the climb component of the Peach-Koehler force $f_{\rm cl}$ and the vacancy concentration in the dislocation core $c^c_d$ depend on $x$: $f_{\rm cl}(x)=\sigma_{33}(x)b$ which is in the inward radial direction and $c^c_d(x)=c^c_0e^{-\frac{\sigma_{33}(x)\Omega}{kT}}$.
Using the climb velocity formula in Eq.~\eqref{eqn:climb-dv0} with arclength $s=R\theta$, and $\frac{\D }{\D s}=\frac{\partial x}{\partial s}\frac{\partial}{\partial x}=-\sin\theta\frac{\partial}{\partial x}$, we have
\begin{equation}
v_{\rm cl}(x,y)=D_c b\frac{\D^2 c^c_d}{\D s^2}
=\frac{D_cb\Omega c^c_d(x)}{kT}\left[-\sin^2\theta\frac{\partial^2 \sigma_{33}}{\partial x^2}
+\frac{\Omega\sin^2\theta}{kT}\left(\frac{\partial \sigma_{33}}{\partial x}\right)^2
+\frac{\cos\theta}{R}\frac{\partial \sigma_{33}}{\partial x}\right].
\end{equation}
This climb velocity depends on both $\frac{\partial \sigma_{33}}{\partial x}$ and $\frac{\partial^2 \sigma_{33}}{\partial x^2}$.

Therefore, our climb velocity formula gives a quantitative understanding of the translation of prismatic dislocation loops due to pipe diffusion. To our knowledge, such a quantitative theory for this phenomenon was not available previously in the literature. { Dislocation dynamics simulations for such translation of prismatic loops under pipe diffusion based on Eqs.~\eqref{eqn:climb-dv0} and \eqref{eqn:climb-dv0-1} are being performed and the results will be reported elsewhere \citep{Niu2016}.}

\section{Conclusions and discussion}

 We have developed vacancy-assisted dislocation climb models from atomistic scheme to dislocation jog dynamics model, and final to dislocation dynamics model on the mesoscopic scale. Our models  incorporate microscopic mechanisms of (i) bulk diffusion of vacancies,
               (ii) vacancy exchange dynamics between bulk and dislocation core,
                (iii) vacancy pipe diffusion along the dislocation core, and
               (iv) vacancy attachment-detachment kinetics at jogs leading  to the motion of jogs.

 The developed microscopic scheme consists of stochastic motion of jogs and hopping  of vacancies at each discrete lattice site.
       By upscaling in space and time from the microscopic model, we obtained the jog dynamics model. This model consists of (1) a vacancy pipe diffusion equation on the jogged dislocation line with boundary condition at each jog depending on the vacancy attachment-detachment kinetics there, (2) a vacancy bulk diffusion equation with boundary condition on the dislocation core surface depending on the vacancy exchange between the bulk and the dislocation core, and (3) a jog velocity formula for the motion of jogs along the dislocation which comes from the inward flux of vacancies and leads to the climb of the dislocation.

We obtained the dislocation dynamics model for dislocation climb by further upscaling in space and time from the jog dynamics model in the fast pipe diffusion limit. Our model consists of the vacancy bulk diffusion equation and a dislocation climb velocity formula. The effects of the above four microscopic mechanisms are incorporated  by a  Robin boundary condition near the dislocations for the bulk diffusion equation
  and a new contribution in the dislocation climb velocity due to vacancy pipe diffusion driven by the stress variation along the dislocation. The latter term is able to quantitatively describe the translation of a prismatic loop at low temperatures when the bulk diffusion is negligible, which was observed in experiments  \citep{Hirth-Lothe,Kroupa1961}.

 Using this new formulation, we derived analytical formulas for the climb velocity of the special cases of a straight edge dislocation and a circular prismatic loop. In the limit cases of negligible stress variation and fast exchange of vacancies between the dislocation core and the bulk, our formulation is reduced to the classical results.
Our formulation can be implemented in three dimensional dislocation dynamics simulations, for example, efficiently using the Green's function or the boundary integral equation method as in \citet{Gu2015319}.

{Real crystalline materials have complicated lattice structures such as fcc, bcc, and hcp than the simple cubic lattice. We have made modifications by using a general dislocation core with radius $r_d$ in our dislocation dynamics and jog dynamics models, so that these models are able to apply to these general lattice structures in the same way as the classical dislocation climb theory \citep{Hirth-Lothe}.  The detailed structures of dislocation cores and jogs in different lattice structures do not change the main features of the above four microscopic mechanisms described by  the climb model on the dislocation dynamics level, in which the influences of these structures  are reflected by values of the parameters in the model. These lattice-dependent parameters can be calculated by atomistic simulations, e.g.~\citet{Hoagland1998,Fang2000,Mishin2009,Yip2009,Yip2010,YZWang2012,Dudarev2016}. Incorporation of more details
of the lattice structures in the dislocation dynamics model for climb  can be further explored in the future work.}

\section*{Acknowledgments}
This work was partially supported by the Hong Kong Research Grants Council General Research Fund 16302115. The work of J.L. was supported in part by National Science Foundation under grant NSF DMS-1454939.


\begin{thebibliography}{29}
\providecommand{\natexlab}[1]{#1}
\providecommand{\url}[1]{\texttt{#1}}
\expandafter\ifx\csname urlstyle\endcsname\relax
  \providecommand{\doi}[1]{doi: #1}\else
  \providecommand{\doi}{doi: \begingroup \urlstyle{rm}\Url}\fi

\bibitem[Arsenlis et~al.(2007)Arsenlis, Cai, Tang, Rhee, Oppelstrup, Hommes,
  Pierce, and Bulatov]{Arsenlis}
A.~Arsenlis, W.~Cai, M.~Tang, M.~Rhee, T.~Oppelstrup, G.~Hommes, T.~G. Pierce,
  and V.~V. Bulatov.
\newblock Enabling strain hardening simulations with dislocation dynamics.
\newblock \emph{Modelling Simul. Mater. Sci. Eng.}, 15:\penalty0 553--595,
  2007.

\bibitem[Ayas et~al.(2012)Ayas, Deshpande, and Geers]{Ayas2012}
C.~Ayas, V.S. Deshpande, and M.G.D. Geers.
\newblock Tensile response of passivated films with climb-assisted dislocation
  glide.
\newblock \emph{J. Mech. Phys. Solids}, 60:\penalty0 1626--1643, 2012.

\bibitem[Ayas et~al.(2014)Ayas, van Dommelen, and Deshpande]{Ayas2014113}
C.~Ayas, J.A.W. van Dommelen, and V.S. Deshpande.
\newblock Climb-enabled discrete dislocation plasticity.
\newblock \emph{J. Mech. Phys. Solids}, 62:\penalty0 113--136, 2014.

\bibitem[Bako et~al.(2011)Bako, Clouet, Dupuy, and Bletry]{Bako}
B.~Bako, E.~Clouet, L.~M. Dupuy, and M.~Bletry.
\newblock Dislocation dynamics simulations with climb: kinetics of dislocation
  loop coarsening controlled by bulk diffusion.
\newblock \emph{Phil. Mag.}, 91:\penalty0 3173--3191, 2011.

\bibitem[Balluffi et~al.(2005)Balluffi, Allen, and Carter]{Balluffi}
R.W. Balluffi, S.M. Allen, and W.C. Carter.
\newblock \emph{Kinetics of Materials}.
\newblock John Wiley \& Sons, Inc., 2005.

\bibitem[Danas and Deshpande(2013)]{Danas2013}
K.~Danas and V.S. Deshpande.
\newblock Plane-strain discrete dislocation plasticity with climb-assisted
  glide motion of dislocations.
\newblock \emph{Modelling Simul. Mater. Sci. Eng.}, 21:\penalty0 045008, 2013.

\bibitem[Fang and Wang(2000)]{Fang2000}
Q.F. Fang and R.~Wang.
\newblock Atomistic simulation of the atomic structure and diffusion within the
  core region of an edge dislocation in aluminum.
\newblock \emph{Phys. Rev. B}, 62:\penalty0 9317--9324, 2000.

\bibitem[Gao et~al.(2011)Gao, Zhuang, Z.L., You, Zhao, and Zhang]{Gao20111055}
Y.~Gao, Z.~Zhuang, Liu Z.L., X.C. You, X.C. Zhao, and Z.H. Zhang.
\newblock Investigations of pipe-diffusion-based dislocation climb by discrete
  dislocation dynamics.
\newblock \emph{Int. J. Plasticity}, 27:\penalty0 1055--1071, 2011.

\bibitem[Geslin et~al.(2015)Geslin, Appolaire, and Finel]{Pierre-Antoine}
P.-A. Geslin, B.~Appolaire, and A.~Finel.
\newblock Multiscale theory of dislocation climb.
\newblock \emph{Phys. Rev. Lett.}, 115:\penalty0 265501, 2015.

\bibitem[Ghoniem et~al.(2000)Ghoniem, Tong, and Sun]{Ghoniem}
N.M. Ghoniem, S.-H. Tong, and L.Z. Sun.
\newblock Parametric dislocation dynamics: a thermodynamics-based approach to
  investigations of mesoscopic plastic deformation.
\newblock \emph{Phys. Rev. B}, 61:\penalty0 913--927, 2000.

\bibitem[Gu et~al.(2015)Gu, Xiang, Quek, and Srolovitz]{Gu2015319}
Y.J. Gu, Y.~Xiang, S.S Quek, and D.J. Srolovitz.
\newblock Three-dimensional formulation of dislocation climb.
\newblock \emph{J. Mech. Phys. Solids}, 83:\penalty0 319--337, 2015.

\bibitem[Gu et~al.(2016)Gu, Xiang, and Srolovitz]{Gu2016}
Y.J. Gu, Y.~Xiang, and D.J. Srolovitz.
\newblock Relaxation of low-angle grain boundary structure by climb of the
  constituent dislocations.
\newblock \emph{Scripta Mater.}, 114:\penalty0 35--40, 2016.

\bibitem[Hafez~Haghighat et~al.(2013)Hafez~Haghighat, Eggeler, and
  Raabe]{Raabe2013}
S.~M. Hafez~Haghighat, G.~Eggeler, and D.~Raabe.
\newblock Effect of climb on dislocation mechanisms and creep rates in
  $\gamma'$-strengthened ni base superalloy single crystal: A discrete
  dislocation dynamics study.
\newblock \emph{Acta Mater.}, 61:\penalty0 3709--3723, 2013.

\bibitem[Hirth and Lothe(1982)]{Hirth-Lothe}
J.P. Hirth and J.~Lothe.
\newblock \emph{Theory of Dislocations}.
\newblock McGraw-Hill, New York, 1982.

\bibitem[Hoagland et~al.(1998)Hoagland, Voter, and Foiles]{Hoagland1998}
R.G. Hoagland, A.F. Voter, and S.M. Foiles.
\newblock Self-diffusion within the cores of a dissociated glide dislocation in
  an fcc solid.
\newblock \emph{Scripta Mater.}, 39:\penalty0 589--596, 1998.

\bibitem[Johnson(1960)]{Johnson1960}
C.A. Johnson.
\newblock The growth of prismatic dislocation loops during annealing.
\newblock \emph{Phil. Mag.}, 5:\penalty0 1255--1265, 1960.

\bibitem[Kabir et~al.(2010)Kabir, Lau, Rodney, Yip, and Van~Vliet]{Yip2010}
M.~Kabir, T.T. Lau, D.~Rodney, S.~Yip, and K.J. Van~Vliet.
\newblock Predicting dislocation climb and creep from explicit atomistic
  details.
\newblock \emph{Phys. Rev. Lett.}, 105:\penalty0 095501, 2010.

\bibitem[Keralavarma et~al.(2012)Keralavarma, Cagin, Arsenlis, and
  Benzerga]{Keralavarma}
S.M. Keralavarma, T.~Cagin, A.~Arsenlis, and A.A. Benzerga.
\newblock Power-law creep from discrete dislocation dynamics.
\newblock \emph{Phys. Rev. Lett.}, 109:\penalty0 265504, 2012.

\bibitem[Kroupa and Prince(1961)]{Kroupa1961}
F.~Kroupa and P.B. Prince.
\newblock Conservative climb of a dislocation loop due to its interaction with
  an edge dislocation.
\newblock \emph{Phil. Mag.}, 6:\penalty0 243--247, 1961.

\bibitem[Lau et~al.(2009)Lau, Lin, Yip, and Van~Vliet]{Yip2009}
T.T. Lau, X.~Lin, S.~Yip, and K.J. Van~Vliet.
\newblock Atomistic examination of the unit processes and vacancy-dislocation
  interaction in dislocation climb.
\newblock \emph{Scripta Mater.}, 60:\penalty0 399--402, 2009.

\bibitem[Lu et~al.(2015)Lu, Liu, and Margetis]{Lu2015}
J.F. Lu, J.-G. Liu, and D.~Margetis.
\newblock Emergence of step flow from an atomistic scheme of epitaxial growth
  in $1+1$ dimensions.
\newblock \emph{Phys. Rev. E}, 91:\penalty0 032403, 2015.

\bibitem[Mordehai et~al.(2008)Mordehai, Clouet, Fivel, and Verdier]{Mordehai}
D.~Mordehai, E.~Clouet, M.~Fivel, and M.~Verdier.
\newblock Introducing dislocation climb by bulk diffusion in discrete
  dislocation dynamics.
\newblock \emph{Philos. Mag.}, 88:\penalty0 899--925, 2008.

\bibitem[Niu et~al.(2016)Niu, Gu, and Xiang]{Niu2016}
X.H. Niu, Y.J. Gu, and Y.~Xiang.
\newblock Conservative climb of prismatic loops by dislocation dynamics
  simulations.
\newblock \emph{In preparation}, 2016.

\bibitem[Pun and Mishin(2009)]{Mishin2009}
G.P. Pun and Y.~Mishin.
\newblock A molecular dynamics study of self-diffusion in the cores of screw
  and edge dislocations in aluminum.
\newblock \emph{Acta Mater.}, 57:\penalty0 5531--5542, 2009.

\bibitem[Sarkar et~al.(2012)Sarkar, Li, Cox, Bitzek, Lenosky, and
  Wang]{YZWang2012}
S.~Sarkar, J.~Li, W.T. Cox, E.~Bitzek, T.J. Lenosky, and Y.Z. Wang.
\newblock Finding activation pathway of coupled displacive-diffusional defect
  processes in atomistics: Dislocation climb in fcc copper.
\newblock \emph{Phys. Rev. B}, 86:\penalty0 014115, 2012.

\bibitem[Swinburne et~al.(2016)Swinburne, Arakawa, Mori, Yasuda, Isshiki,
  Mimura, Uchikoshi, and Dudarev]{Dudarev2016}
T.D. Swinburne, K.~Arakawa, H.~Mori, H.~Yasuda, M.~Isshiki, K.~Mimura,
  M.~Uchikoshi, and S.L. Dudarev.
\newblock Fast, vacancy-free climb of prismatic dislocation loops in bcc
  metals.
\newblock \emph{Scientific Report}, 6:\penalty0 30596, 2016.

\bibitem[Turnbull(1970)]{Turnbull1970}
J.A. Turnbull.
\newblock The coalescence of dislocation loops by self climb.
\newblock \emph{Phil. Mag.}, 21:\penalty0 83--94, 1970.

\bibitem[Xiang and Srolovitz(2006)]{Xiang2006}
Y.~Xiang and D.J. Srolovitz.
\newblock Dislocation climb effects on particle bypass mechanisms.
\newblock \emph{Philos. Mag.}, 86:\penalty0 3937--3957, 2006.

\bibitem[Xiang et~al.(2003)Xiang, Cheng, Srolovitz, and E]{Xiang20035499}
Y.~Xiang, L.-T. Cheng, D.J. Srolovitz, and W.~E.
\newblock A level set method for dislocation dynamics.
\newblock \emph{Acta Mater.}, 51:\penalty0 5499--5518, 2003.

\end{thebibliography}
\end{document}